 \documentclass[iop]{emulateapj}
\usepackage{natbib}
\usepackage{amsmath}
\usepackage{dsfont}
\usepackage{snapshot}

\newcommand{\frhat}{\bigodot_L}

\shorttitle{Structure of diffuse emission in Hi-GAL}
\shortauthors{Elia et al.}

\begin{document}


\title{Characterizing the structure of diffuse emission in Hi-GAL maps}


\author{D. Elia\altaffilmark{1}, F. Strafella\altaffilmark{2}, N. Schneider\altaffilmark{3}, R. Paladini\altaffilmark{4}, R. Vavrek\altaffilmark{5},  Y. Maruccia\altaffilmark{2}, S. Molinari\altaffilmark{1}, S. Pezzuto\altaffilmark{1}, A. Noriega-Crespo\altaffilmark{5}, K. L. J. Rygl\altaffilmark{6}, A. Di Giorgio\altaffilmark{1}, A. Traficante\altaffilmark{7}, E. Schisano\altaffilmark{5}, L. Calzoletti\altaffilmark{8}, M. Pestalozzi\altaffilmark{1}, J. S. Liu\altaffilmark{1}, P. Natoli\altaffilmark{9,10,8,11}, M. Huang\altaffilmark{12}, P. Martin\altaffilmark{13}, Y. Fukui\altaffilmark{14}, T. Hayakawa\altaffilmark{14}}
%, J. G. Ingalls\altaffilmark{4}
\altaffiltext{1}{IAPS-INAF, Via Fosso del Cavaliere 100, 00133 Roma, Italy}
\altaffiltext{2}{Dipartimento di Fisica, Universit\`{a} del Salento, CP 193, 73100 Lecce, Italy}
\altaffiltext{3}{IRFU/SAp CEA/DSM, Laboratoire AIM CNRS - Universit\'{e} Paris Diderot, 91191 Gif-sur-Yvette, France}
\altaffiltext{4}{Infrared Processing and Analysis Center, California Institute of Technology, Pasadena, CA 91125, USA}
\altaffiltext{5}{Herschel Science Centre, European Space Astronomy Centre, Villafranca del Castillo. Apartado de Correos 78, 28080 Madrid, Spain}
\altaffiltext{6}{Research and Scientific Support Department, European Space Agency (ESA-ESTEC), PO Box 299, 2200 AG, Noordwijk, The Netherlands}
\altaffiltext{7}{Jodrell Bank Centre for Astrophysics, School of Physics and Astronomy, University of Manchester, Manchester, M13 9PL, UK}
\altaffiltext{8}{Agenzia Spaziale Italiana Science Data Center, c/o ESRIN, via Galileo Galilei, 00044 Frascati, Italy}
\altaffiltext{9}{Dipartimento di Fisica e Science della Terra, Universit\`{a} di Ferrara Via Saragat, 1, 44100 Ferrara, Italy}
\altaffiltext{10}{INFN, Sezione di Ferrara, via Saragat 1, 44100 Ferrara, Italy}
\altaffiltext{11}{INAF-IASF Bologna, Via P. Gobetti 101, 40129, Bologna}
\altaffiltext{12}{National Astronomical Observatories, Chinese Academy of Sciences, Beijing 100012, China}
\altaffiltext{13}{Canadian Institute for Theoretical Astrophysics, University of Toronto, 60 St. George Street, Toronto, ON M5S 3H8, Canada}
\altaffiltext{14}{Department of Physics, Nagoya University, Furo-cho, Chikusa-ku, Nagoya 464-8602, Japan}

%%%%
%%%%\author{D. Elia\altaffilmark{1}, et al.}
%%%%\altaffiltext{1}{IAPS-INAF, Fosso del Cavaliere 100, 00133 Roma, Italy}

%\affil{Astronomy Department, University of California,
%    Berkeley, CA 94720}

%\author{C. D. Biemesderfer\altaffilmark{4,5}}
%\affil{National Optical Astronomy Observatories, Tucson, AZ 85719}
\email{davide.elia@iaps.inaf.it}

%\and
%
%\author{R. J. Hanisch\altaffilmark{5}}
%\affil{Space Telescope Science Institute, Baltimore, MD 21218}


%%\altaffiltext{1}{Visiting Astronomer, Cerro Tololo Inter-American Observatory.
%%CTIO is operated by AURA, Inc.\ under contract to the National Science
%%Foundation.}
%%\altaffiltext{2}{Society of Fellows, Harvard University.}
%%\altaffiltext{3}{present address: Center for Astrophysics,
%%    60 Garden Street, Cambridge, MA 02138}
%%\altaffiltext{4}{Visiting Programmer, Space Telescope Science Institute}
%%\altaffiltext{5}{Patron, Alonso's Bar and Grill}


\defcitealias{eli13}{Paper~I}

\begin{abstract}
We present a study of the structure of the Galactic interstellar medium through the $\Delta$-variance
technique, related to the power spectrum and the fractal properties of infrared/sub-mm maps. Through
this method, it is possible to provide quantitative parameters which are useful to characterize different
morphological and physical conditions, and to better constrain the theoretical models.
In this respect, the \emph{Herschel}\footnote{\emph{Herschel} is an ESA space observatory with science
instruments provided by European-led Principal Investigator consortia and with important participation
from NASA.} Infrared Galactic Plane Survey carried out at five photometric
bands from 70 to 500~$\mu$m constitutes an unique database for applying statistical tools to a variety
of regions across the Milky Way. In this paper, we derive a robust
estimate of the power-law portion of the power spectrum of four contiguous $2^{\circ} \times 2^{\circ}$
Hi-GAL tiles located in the third Galactic quadrant
($217^{\circ} \lesssim \ell \lesssim 225^{\circ}$, $-2^{\circ}\lesssim b \lesssim 0^{\circ}$). The low
level of confusion along the line of sight testified by CO observations makes this region an ideal case.
We find very different values of the power spectrum slope from tile to tile but also from wavelength to
wavelength ($2 \lesssim \beta \lesssim 3$), with similarities between fields attributable to components
located at the same distance. Thanks to the comparison with models of turbulence, an
explanation of the determined slopes in terms of the fractal geometry is also provided,
and possible relations with the underlying physics are investigated. In particular, an
anti-correlation between ISM fractal dimension and star formation efficiency
is found for the two main distance components observed in these fields. A possible link
between the fractal properties of the diffuse emission and the resulting clump mass
function is discussed.

\end{abstract}





\section{Introduction}\label{intro}

One of the most intriguing tasks in the observational study of the
interstellar medium (ISM) is to extract information about the
3-dimensional structure of the clouds, starting from the
2-dimensional maps of these objects, generally taken at different
wavelengths and with different techniques and resolutions. Although
a certain degree of self-similarity of the ISM maps over a given range
of spatial scales can be in many cases perceived by eye,
there are numerous more solid arguments suggesting this can be the case,
starting from the work of \citet{sca90}.

In this respect the phenomenon mainly responsible of self-similar morphologies
is turbulence. This is a largely recognized fact in molecular clouds, being
a typical scale-free phenomenon inducing fractality \citep[see, e.g.,][]{sre89}. It
is indeed characterized by the lack of a specific length scale, then it can produce
a fractal distribution of matter in a molecular cloud over a wide range of scales.
Therefore, to determine the starting and the ending point of these ranges is generally
considered a tentative way to get an estimate of the turbulence injection and
dissipation scales. An extensive and detailed review of the observational evidences
of the presence of turbulence in molecular clouds and its role in shaping their
structure in fractal sense can be found in \citet{vaz99} and \citet{sch11}.

It is noteworthy that the ISM clouds belong to the category of the
\emph{stochastic fractals}, whose structure does not appear perfectly
self similar, but rather \emph{self-affine}: although a given set and a
part of it have not exactly the same appearance, they have the same statistical
properties and it is still possible to use a fractal description for them.

There are many observational grounds supporting the fractal scenario. The
observations of the low-$J$ $^{12}$CO and $^{13}$CO emission lines in several
star-forming molecular cloud complexes \citep[e.g.,][]{fal96,sch98,wil99b}
show that the measured line intensities, shapes and ratios cannot be
produced in clouds of uniform gas temperature and density, suggesting
the idea that these interstellar objects are far from being homogeneous,
being instead organized in small clumps with a filling factor lower than
unity \citep{elm97a}. Interestingly, such a structure is also
able to justify further observed characteristics of the investigated
region, as for example the clump mass function \citep{sha11} and
the stellar initial mass function \citep{elm02}. These remain meaningful
observables although in the last years the picture of the ISM has changed
with the recognition of filaments as intermediate structures
\citep[e.g.,][]{ros96,wil99b}, which have definitely been found ubiquitous
in the recent Herschel observations \citep[e.g.,][]{mol10b,sch13}. In any
case, the cloud description based on a hierarchical decomposition in
recognizable substructures \citep{hou92} is not incompatible with the
fractal approach. Indeed \citet{stu98} have shown that these two points
of view are consistent: an ensemble of clumps with a given mass and
size spectrum can give rise to a fractal structure of the cloud.

Statistical descriptors which, in general, can be related to the fractal
properties of a cloud are powerful methods to characterize its structure.
The techniques initially used to estimate the fractal dimension of the
interstellar clouds were based on the isocontours of the images, as
for example the \emph{perimeter-ruler} and the \emph{area-perimeter}
relation \citep[see e.g.][and references therein]{san05}. Subsequently,
statistical tools have been applied, namely descriptors based on the
value and the spatial distribution of the single pixels, providing
quantitative information on one or more aspects of the investigated
morphology \citep[a relevant part of them is summarized in][]{elm04}. The
direct estimate of the power spectrum \citep[e.g.][]{ing04,miv07,mar10,gaz10}
can be used to infer the fractal structure of the ISM, although to deal
with real observational sets other algorithms have been demonstrated to
be more adequate \citep{stu98}. Other statistical estimators are the structure
function \citep{padc02,pad03,kri04,cam05,gus06,kow07,row11}, the $\Delta$-variance
(see below), the autocorrelation function \citep{cam05}, and the adapted
correlation length \citep{car06}, whereas a further development of these
monofractal descriptors is represented by the multifractal spectrum
\citep{cha01,vav01}.
In particular, the $\Delta$-variance method was introduced by \citet{stu98}
and subsequently improved by \citet{ben01} and \citet{oss08a} to analyze
the drift behavior of observed scalar functions such as the intensity distribution
in molecular clouds, real or synthesized. It has been applied not only to maps
of line emission \citep[see also][]{ben01,oss01,oss08a,sch11,row11}
and dust extinction \citep{cam04,sch11} or emission \citep{rus13}, but also to
the recovered velocity field \citep{oss06,fed10}, or to 3-dimensional density
fields of turbulence simulations \citep{fed09}.

The aim of this paper is to contribute both to the enlargement of the
sample of the regions whose structural properties have been studied by
means of fractal techniques, and to the improvement in characterizing
the response of statistical tools to different observing conditions. The
Hi-GAL survey \citep[\emph{Herschel} Infrared GALactic plane survey,][]{mol10a}
represents an extraordinary resource for carrying out statistical studies
of the ISM. Indeed a large coverage is obtained in five different bands,
so that a large variety of morphologies and physical conditions can be
investigated at unprecedented spatial resolution. Moreover, these large
\emph{Herschel} maps offer the chance to probe a wide range of spatial scales,
since the number of available pixels is very important for the
reliability of the statistical descriptors.

Galactic plane observations suffer from confusion due to the superposition
of different components along the line of sight, especially in the first
and fourth Galactic quadrants. To minimize the problem of confusion, the
first available observations of the third Galactic quadrant
(in the range $217.0^{\circ} \lesssim \ell \lesssim 224.3^{\circ}$) are
studied as a first test case, in which we are more confident that the
observed ISM emission corresponds to a morphology which is quite coherent
from the spatial point of view. These observations have been presented by
\citet{eli13} \citepalias[hereafter][]{eli13}, and are briefly summarized
in Section~\ref{obs}.

As a paradigm of synthetic cloud images used for testing the statistical
tools used in this work we consider the class of so-called \emph{fractional
Brownian motion} images (hereinafter fBm). They have been already used,
for example, by \citet{stu98,ben01,kha06,miv07,sha11} for testing their
algorithms. We briefly discuss the properties of this class of images
in Section~\ref{fbm_par}.

In this paper we adopt the $\Delta$-variance algorithm to derive
a robust estimate of the power spectrum slope of the maps. In
Section~\ref{dvar_par} this method is briefly described, and
its application to synthetic maps is discussed to characterize
the response of the algorithm in case of departure of the analyzed
image from the ideal fBm-like behavior.

In Section~\ref{dvarhigal_par} we preent the results of our $\Delta$-variance
analysis, and discuss the obtained power spectrum slopes and self-similarity
ranges, searching for cross-correlations among different maps and observational
wavebands. Moreover, links with turbulence and observables related to star
formation (as star formation efficiency and mass functions) are investigated.
Finally, the results are summarized in Section~\ref{summary}.

\section{Observational datasets}\label{obs}

The \emph{Herschel} \citep{pil10} open time key project Hi-GAL \citep{mol10a} is
a five-band photometric survey initially aimed at studying the stellar life cycle
in the inner Galaxy ($-72^{\circ} \lesssim \ell \lesssim 68^{\circ}$) and
subsequently extended to the whole Galactic plane.

The first Hi-GAL observations available for the outer Galaxy, presented in
\citetalias{eli13}, consist of four $2.3^{\circ} \times 2.3^{\circ}$ adjacent tiles
of the Galactic third quadrant. We will denote these with $\ell217$, $\ell220$,
$\ell222$, $\ell224$, respectively, according to the \emph{Herschel} Data Archive
nomenclature. These far-infrared maps of the outer Galaxy represent an ideal case
for studying the structure of the ISM, for two main reasons: the lower occurrence,
in general, of compact bright sources and of star forming regions, and the lower
degree of confusion along the line of sight.

The reduction procedure and the main characteristics of these observations are
described in \citetalias{eli13}, so here we resume only those features which
are useful for the discussion in this paper. The observed wavebands are
centered around 70 and 160~$\mu$m \citep[PACS,][]{pog10} and 250, 350 and 500~$\mu$m
\citep[SPIRE,][]{gri10}, with nominal resolutions of about $5 \arcsec$,
$12 \arcsec$, $18 \arcsec$, $25 \arcsec$, $36 \arcsec$, respectively.
The pixel sizes of these maps are $3.2 \arcsec$, $4.5 \arcsec$,
$6.0 \arcsec$, $8.0 \arcsec$, $11.5 \arcsec$, at  70, 160, 250, 350 and
500~$\mu$m, respectively.
%Observing strategy and data reduction for the Hi-GAL maps are
%described in detail in \citet{mol10b} and \citet{tra11}.
The fields were observed by simultaneously acquiring PACS and SPIRE images
in the five aforementioned photometric bands. This observing mode generally
implies that the areas imaged by the two instruments are not exactly the same;
in this paper, in particular, we consider only the common area of each tile,
because we are interested in comparing the statistics of the same regions of
the sky seen at different wavelengths.
%Further astrometric registration of possible small coordinate offsets has
%been performed, based on the comparison between the positions of bright isolated
%70~$\mu$m sources and of their possible counterparts in the WISE point source catalog
%\citep{wri10} endowed with a reliable flux at $22 \mu$m. Furthermore,

\begin{figure*}[p]
\begin{minipage}{15.7cm}\begin{center}\plotone{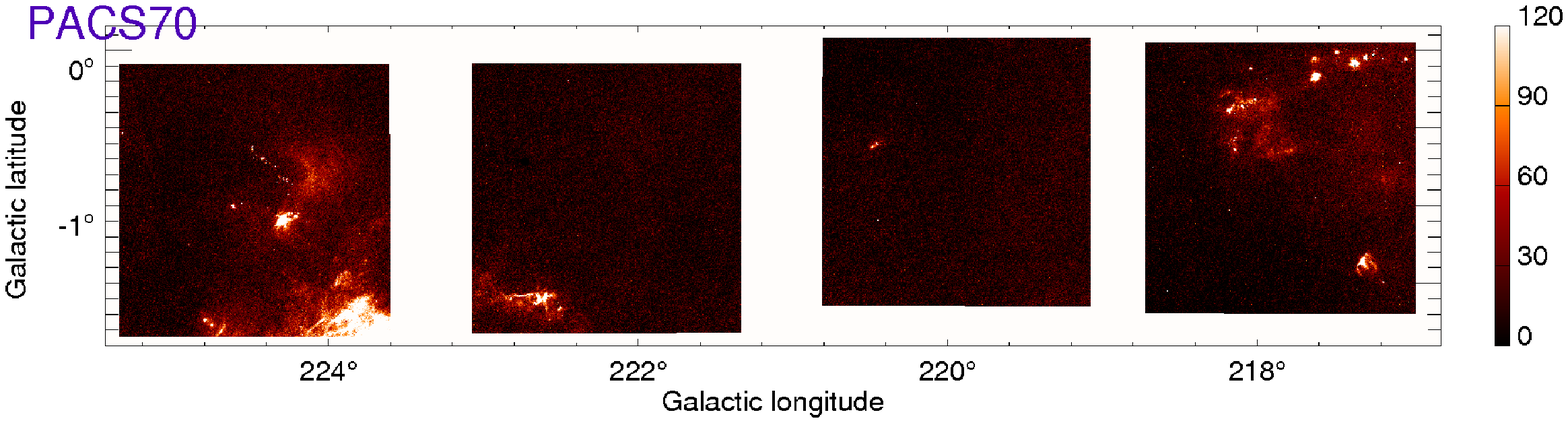}\end{center}\end{minipage}
\begin{minipage}{15.7cm}\begin{center}\plotone{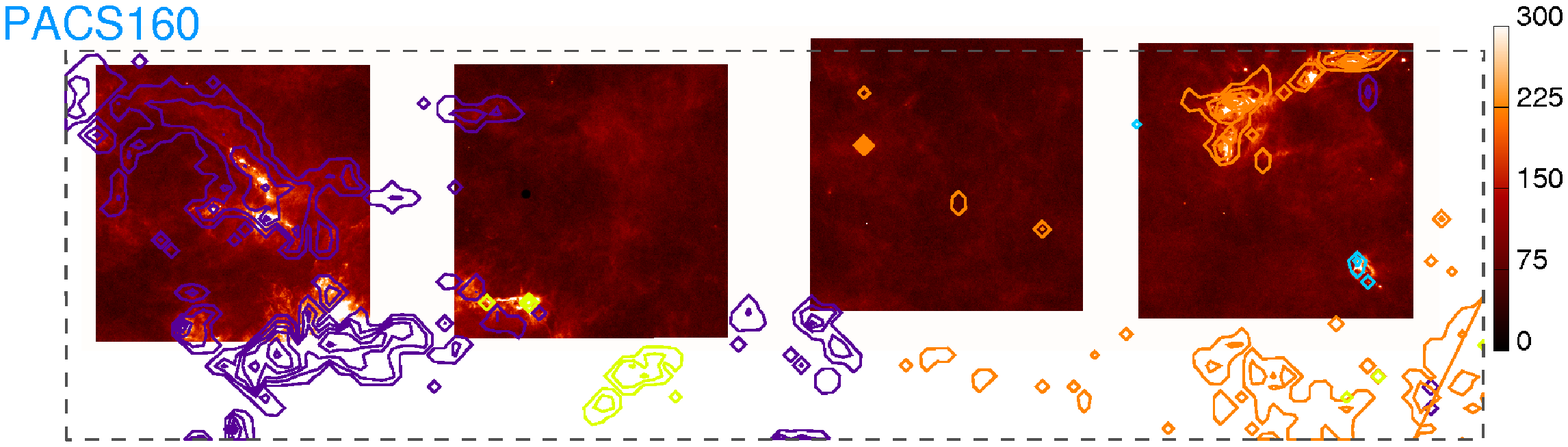}\end{center}\end{minipage}
\begin{minipage}{15.7cm}\begin{center}\plotone{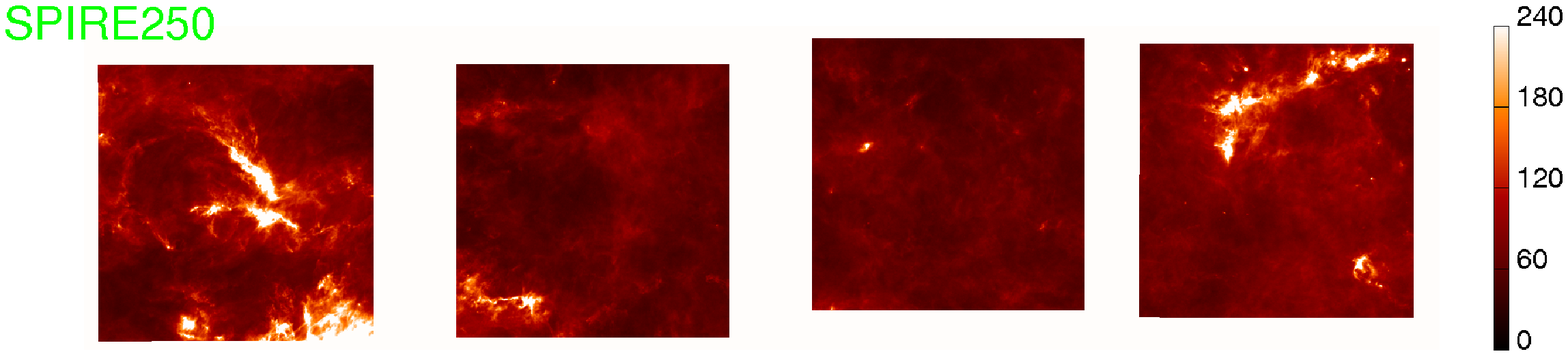}\end{center}\end{minipage}
\begin{minipage}{15.7cm}\begin{center}\plotone{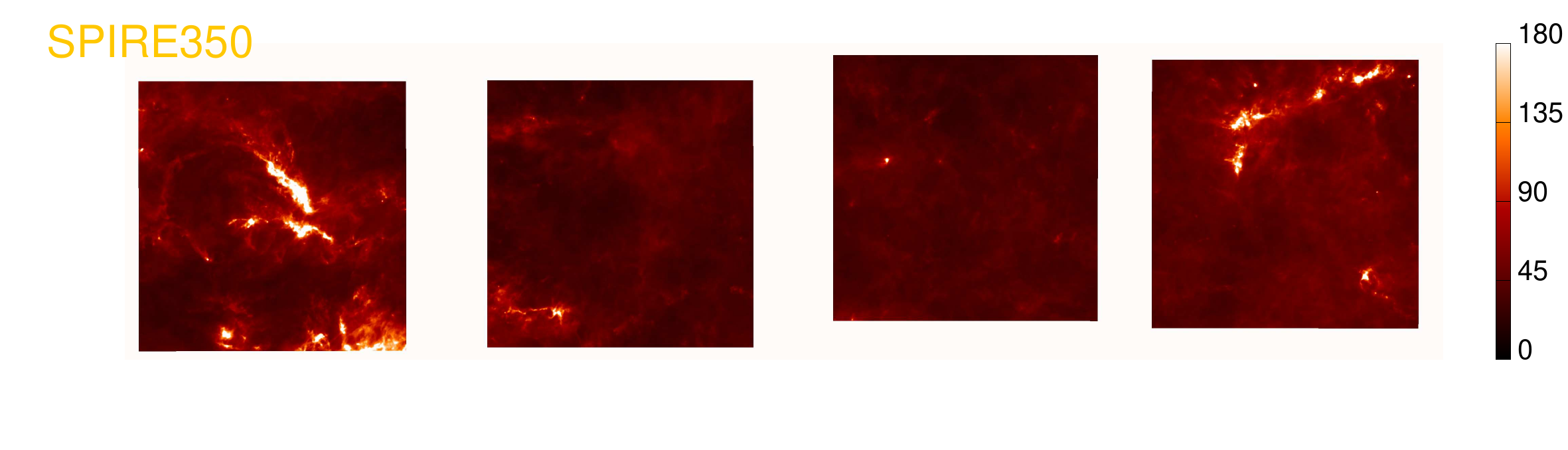}\end{center}\end{minipage}
\begin{minipage}{15.7cm}\begin{center}\plotone{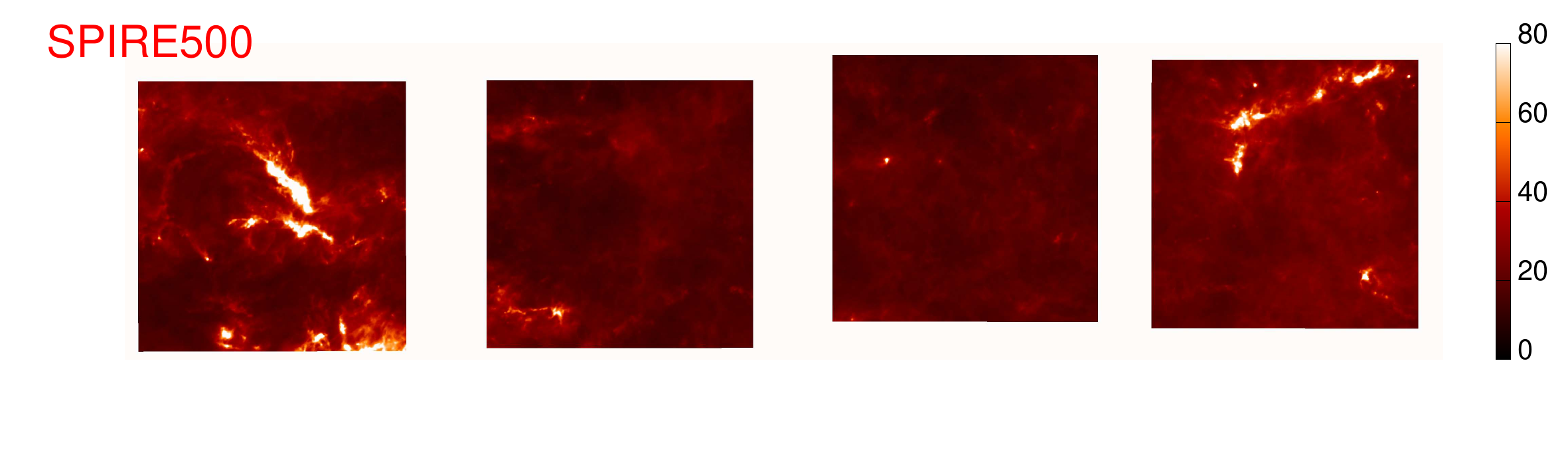}\end{center}\end{minipage}
\begin{minipage}{15.7cm}\begin{center}\plotone{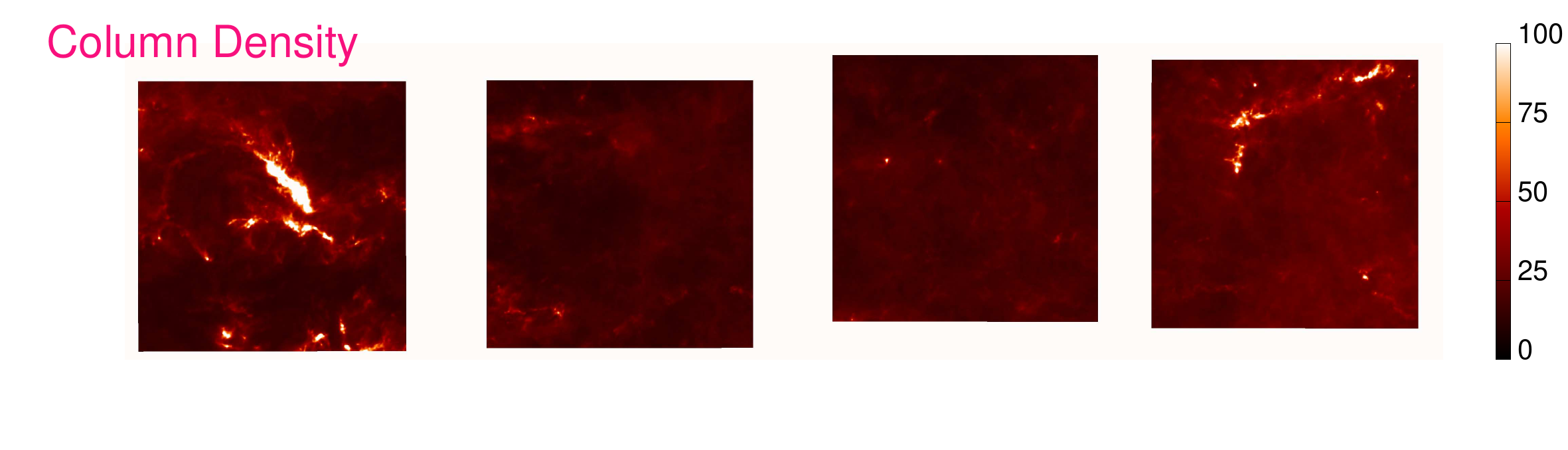}\end{center}\end{minipage}
\caption{Fields investigated in this paper, divided by observed band. The
instrument/wavelength combination is specified in each raw (the column density
maps are also shown in the last raw), adopting a color convention which is used
throughout the rest of the paper (blue: 70~$\mu$m, cyan: 160~$\mu$m, green:
250~$\mu$m, orange: 350~$\mu$m, red: 500~$\mu$m, magenta: column density).
The color scales are linear, and the top value (corresponding to the white level)
is chosen to allow a satisfactory visualization of the lower emission levels;
the units are MJy/sr for the PACS and SPIRE maps, and $10^{20}$~cm$^{-2}$ for
the column density ones. The Galactic coordinate grid is displayed in the
70~$\mu$m case to show the displacement in the sky of the four considered tiles.
The CO(1-0) contour levels \citepalias{eli13} are overplotted on the map at 160~$\mu$m.
They start from 5 K km $s^{-1}$ and are in steps of 15 K km $s^{-1}$. Components~I, II,
III and IV are represented with purple, orange, yellow and cyan contours, respectively.
The area surveyed in CO(1-0) is delimited by a grey dashed line.
\label{higal2}}
\end{figure*}

The final images extracted at each wavelength for each tile are shown in
Figure~\ref{higal2}, where the color coding used to identify the different
\emph{Herschel} bands throughout the article is introduced. Since for a given tile the
same area of the sky is considered in each band, the total number of pixels depends
on the band: the size of each image in pixels is reported in Table~\ref{imasizes}.

\begin{deluxetable}{lccccc}
\tabletypesize{\scriptsize}
\tablecaption{Sizes in pixels of the investigated images\label{imasizes}}
\tablewidth{0pt}
\tablehead{
\colhead{Tile} & \colhead{70 $\mu$m} & \colhead{160 $\mu$m} & \colhead{250 $\mu$m} & \colhead{350 $\mu$m} &
\colhead{500 $\mu$m}\tablenotemark{a}}
\tablenotetext{a}{The size of the column density maps is the same as that of the corresponding 500 $\mu$m maps.}
\startdata
$\ell$217 & 1971 & 1403 & 1051 & 789 & 549 \\
$\ell$220 & 1951 & 1387 & 1041 & 781 & 545 \\
$\ell$222 & 1961 & 1395 & 1047 & 785 & 547 \\
$\ell$224 & 1977 & 1405 & 1055 & 791 & 551 \\
\enddata
\end{deluxetable}

The column density maps of each tile are analyzed here as well. They have been
derived from a pixel-to-pixel modified black body fit as explained in more detail
in \citetalias{eli13}. To this end, first the maps were  absolutely calibrated
correcting their zero level by means of offset values derived from the comparison
with the Planck/IRAS data \citet{ber10}. Then, the 160, 250, and 350~$\mu$m maps were
reprojected onto the grid of the 500~$\mu$m ones. The 70~$\mu$m maps were not involved
in this calculation because they can contain emission coming from the so-called
Very Small Grains \citep{com10}, not at thermal equilibrium, and/or from warmer spectral
components, such as the proto-stellar content of the clumps, reflection nebulae, etc.
For these reasons the column density maps we obtained are more suitable to describe the
cold component of dust in these regions. At a first glance, they look very similar
to the SPIRE maps, and in particular to the 500~$\mu$m ones, having also the same
resolution: both these aspects turn out to be important for the spatial analysis
reported in the following sections.

The four tiles show the presence of star forming activity, but $\ell$217 and
$\ell$224 contain the brightest and most extended regions, mainly arranged in
large filamentary shapes.

The knowledge of the kinematic distances of the clouds is somehow fundamental
for identifying really coherent nebular structures, instead of superpositions of
different distance components, to which the statistical descriptors can be applied.
The distances for the Hi-GAL tiles analyzed here are obtained
from CO(1-0) line observations carried out at the NANTEN sub-mm telescope
\citep{miz04}, and presented in detail in \citetalias{eli13}. In detail, gas
emission in the $\ell$217 field is dominated by a component located at a kinematic
distance of $\sim 2.2$~kpc \citepalias[see Table~1 of][their component~II]{eli13}.
CO emission is fainter in the $\ell$220 tile, yet the component corresponding to
the 2.2~kpc distance is still predominant.

Looking at the second raw of Figure~\ref{higal2}, the $\ell$222 and $\ell$224
tiles are found to be dominated by a bright component
corresponding to an average distance of $\sim 1.1$~kpc, i.e. the so-called
component~I of \citetalias{eli13}. Two further distance components, III and IV, can
be considered negligible in the present analysis. The former, located at an average
distance of 3.3~kpc (then likely in the Perseus arm) coincides with a bright region in
the south-eastern corner of the original $\ell$222, which however lies outside the region
considered in this paper as a consequence of the tile cropping described above. The
latter is constituted only by a bright but small portion of the $\ell$217 field.

\section{Power spectrum and fBm images}\label{fbm_par}

The statistical descriptors we use in this paper revolve around the central concept
of \emph{power spectrum} $\mathcal{P}(\mathbf{k})$ of the observable $A(\mathbf{x})$,
defined as the square modulus of its Fourier transform $\tilde{A}(\mathbf{k})$.
The variable $\mathbf{x}$ is defined in a $E$-dimensional space, with $E=2$ in
the case of image analysis. A convenient choice is to study the shell-averaged
power spectrum $\mathcal{P}(k)$, where $k=\|\mathbf{k}\|^2$.
Because a power law is a distinctive experimental signature seen in a large variety
of complex systems, frequently a search for a power-law behavior of the spectrum is
carried out. An emblematic case is that of the Kolmogorov's power spectrum
of turbulence \citep{kol41}, calculated for the velocity field of incompressible
fluids; a dependence as $\mathcal{P}(k)\propto k^p$ is expected, with $p=-11/3$
in three dimensions, $p=-8/3$ in two and $p=-5/3$ in one, respectively. Different
models, as for example the classic Burgers' turbulence \citep[e.g.][]{bec07}, still
show a similar power spectrum, although with different slope \citep[$p=2$, ][]{bis03}.
The presence of such a power-law behavior (in the full range of values
of the $k$ wave number, or over a limited part of it) can be interpreted
as an indication of turbulence and can suggest the characteristic
scales at which both energy injection and dissipation take place.

The power spectra (or portions of them) of the ISM maps often exhibit a
power-law behavior. The link with turbulence is quite natural: it is commonly
accepted that a turbulent velocity field in the ISM can also shape the
density field \citep[cf.][]{bol02,pad03,row11}. To derive the slope of
the power spectrum, however, requires some care. Indeed, the Fourier
transform of a non-infinite mapped signal inevitably introduces unwanted
frequencies due to the spatial sampling and to the limited size of the image,
since the Fourier transform implicitly assumes wrap-around periodicity) leading to
aliasing \citep{stu98,ben01}. This has lead, in the past, to the use of
statistical tools more robust than the direct determination of power
spectrum. The improved performance of the new generation of observing
instruments have allowed the production of maps with large numbers
of pixels (as in the case of the Hi-GAL tiles), which might mitigate
the aforementioned issues. Nevertheless in this paper we prefer to keep
exploiting one of these indirect methods, namely the $\Delta$-variance
technique, briefly discussed in Section~\ref{dvar_par}. In this way we
can both make possible a direct comparison with the literature, and
exploit further information that this technique can provide about the
structure of the maps (see Section~\ref{dvar_par}). % and  and \ref{sf_par}

\subsection{Fractional Brownian motion images}\label{fbm_sub}

\begin{figure}[t]
\epsscale{1.0}\plotone{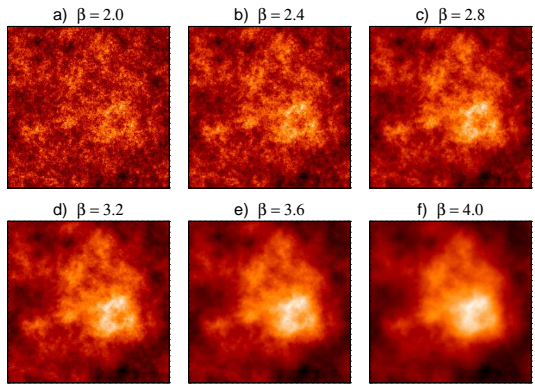} \caption{Synthetically generated fBm
$300\times 300$ pixels images generated with increasing power spectrum
slope $\beta$ starting from the same random phase distribution. Map units
are arbitrary.\label{fbms}}
\end{figure}

There is a category of stochastic fractals that can be helpful to test
the statistical algorithms for structure analysis, since their Fourier transform
has specific analytic properties which make them easy to be generated: the
\emph{fractional Brownian motion} \citep{pei88}. In the two-dimensional
case, these show a good similarity with molecular cloud maps
\citep[][see also Figure~\ref{fbms}]{stu98,ben01,miv03}. A detailed
description of the fBm image properties is given in \citet{stu98};
here we summarize only the most relevant ones, which will prove to be
useful in the analogy with ISM maps we will make in next sections.
First, they are characterized by having a power-law power spectrum
with exponent $\beta=E+2H$, where $E$ is the Euclidean dimension
of the considered space (for cloud maps, $E=2$), and $H$ is the
so-called \emph{Hurst exponent}, ranging from 0 to 1. Therefore
$\beta$ can take values from 2 to 4. Second, the phases of their
Fourier transform are random. Based on these two constraints,
it is quite easy to obtain fBm images once $\beta$ is assigned and
a random phase distribution is generated\footnote[1]{The fBm
images generated in such a way are periodic, i.e. it is
possible to place side by side and to connect with continuity
the image to itself. Clearly, this can not be the case of real
ISM maps.}. In Figure~\ref{fbms}, six $300 \times 300$ pixel
fBm images are shown. They have the same phase distribution and differ
only by $\beta$, which ranges from 2.0 to 4.0 in steps of 0.4. It can be
seen that the phase distribution determines the overall appearance of
the ``cloud'', but as $\beta$ increases the structure becomes smoother
and smoother due to the transfer of power from high to low spatial
frequencies.We recall that the case $\beta=0$ would correspond
to \emph{white noise}.

In terms of the fractal description it has been shown that the fractal
dimension of an $E$-dimensional
fBm image is given by:
\begin{equation}
\label{dh} D=E+1-H\:,
\end{equation}
so that the direct relation between $D$ and $\beta$ is:
\begin{equation}\label{deltafrac}
D=\frac{3E+2-\beta}{2}\:.
\end{equation}

An important property of the fBm images is that the power spectrum of
the ($E-1$)-dimensional projection on an $E$-dimensional fBm set is
again a power-law with the same spectral index \citep{stu98}. This
result turns out to be important for establishing a link between
the observed 2-dimensional column density and the real 3-dimensional
cloud density field for clouds that can be considered isotropic, as
suggested through theoretical arguments \citep{bru10}
and empirical evidences \citep{fed09,gaz10}.

Thus, invoking Equation~\ref{deltafrac}, it is found that
\begin{equation}\label{deltadiff}
D_{E-1}=D_{E}-\frac{3}{2}\,, 
\end{equation}
i.e.~the fractal dimension of a fBm object changes, under projection, 
by 1.5, and not by 1 as one could erroneously expect.

%Thanks to this result, when a two-dimensional fBm image is considered
%as a projection of a fractal ``cloud'' embedded in the three-dimensional
%space, one can determine the spectral index $\beta$ of the
%two-dimensional image, consider the same for the ``cloud'' and derive
%the fractal dimension of this latter using the
%Equation~\ref{deltafrac} for $E=3$. For $2<\beta<4$ we obtain
%$2<D<3$, as expected for a fractal surface embedded in the
%three-dimensional Euclidean space.

\section{$\Delta$-variance analysis}\label{dvar_par}
The $\Delta$-variance method is a generalization of the {\em Allan
variance} \citep{all66}, elaborated and characterized in detail
by \citet{stu98}. For a two-dimensional observable field $A(x,y)$,
the $\Delta$-variance at the scale $L$ is defined as the
variance of the convolution of $A$ with a filter function $\frhat$:
\begin{equation}\label{dvar_eq}
\sigma_{\Delta}^2(L)=\frac{1}{2\pi}\left<\left(A * \frhat
\right)^2\right>_{x,y} \:,
\end{equation}
where
\begin{equation}\label{frhat}
\frhat(r)=\begin{cases}
\frac{1} {\pi (L/2)^2}  & (r \leq \frac{L}{2})                \\
-\frac{1}{8\pi (L/2)^2} & (\frac{L}{2} < r \leq \frac{3L}{2}) \\
0                       & (r > \frac{3L}{2})
\end {cases}
\end{equation}
is the {\em down-up-down cylinder} (or {\em French hat}) function
and $r=\sqrt{x^2+y^2}$. The two non-zero terms of the above definition
represent the \emph{core} and the \emph{annulus} component of the filter,
respectively.

\citet{oss08a} recommended, as a possible alternative to the French
hat function to obtain a more reliable estimate of the spectral index
$\beta$, to use the smoother {\em Mexican hat}, defined as
\begin{equation}
%\mxhat(r)
\frhat(r)=\frac{4}{\pi L^2} \, e^{\frac{r^2}{(L/2)^2}}-\frac{4}{\pi L^2 (v^2-1)}\, \left[e^{\frac{r^2}{(vL/2)^2}} - e^{\frac{r^2}{(L/2)^2}}\right] \;,
\end{equation}
where the two main terms in the right side of the equation represent the
\emph{core} and the \emph{annulus} components, respectively, and $v$ is the
diameter ratio between them.  To speed up calculations, the same
authors suggest to perform the operation in Equation~\ref{dvar_eq}
as a multiplication in the Fourier domain,
\begin{equation}\label{dvar_fourier}
\sigma_{\Delta}^2(L)=\frac{1}{2\pi}\int\int \mathcal{P}
\left|\tilde{\frhat} \right|^2 dk_x dk_y \:,
\end{equation}
where $\mathcal{P}$ is the power spectrum of $A$, and
$\tilde{\frhat}$ is the Fourier transform of the filter function.

The fundamental relation that relates the slopes of $\Delta$-variance
and of the power spectrum ($\beta$) was shown by \citet{stu98}:
\begin{equation}\label{dvar_beta}
\sigma_{\Delta}^2(L) \propto L^{\beta-2}\:.
\end{equation}
Given the expression above, one can derive the power spectrum slope
by performing a linear fit over the range of spatial scales for which
the logarithm of $\Delta$-variance manifests a linear behavior. In
this work, we adopted this procedure following the prescriptions
of \citet{oss08a}\footnote[2]{The IDL package for calculating the
$\Delta$-variance can be found at
\url[http://hera.ph1.uni-koeln.de/~ossk/Myself/deltavariance.html]{http://hera.ph1.uni-koeln.de/$\sim$ossk/Myself/deltavariance.html}.},
i.e. using a Mexican Hat filter with $v=1.5$. Furthermore, we do not
adopt any strategy based on assigning different weights to the pixels
involved in the $\Delta$-variance calculation, which is recommended by
\citet{oss08a} in case of maps characterized by a variable data reliability.
The portions of the Hi-GAL maps we chose, indeed, being far from the
tile boundaries, are characterized by a quite uniform coverage
\citep[see also][]{tra11} and, consequently, by a stable rms noise.

\subsection{The contribution of compact sources}\label{compact}
From Equation~\ref{dvar_beta} the $\Delta$-variance behavior of a fBm
image is expected to be a perfect power law. Although the ISM maps
generally exhibit a fBm-like behavior (see Sect.~\ref{fbm_sub}), it is
important to identify and characterize all the signatures in the power
spectrum ascribable to possible departures from the ideal fBm case, first of
all the presence of bright compact sources. For this purpose, we performed a
test by simulating a PACS $160~\mu$m map of a portion of the Galactic plane,
using some typical parameters of this band, such as the pixel size of $4.5\arcsec$.
The steps of the recipe can be also followed in Figure~\ref{dvarsimul},
together with the effects they achieve on the corresponding $\Delta$-variance
curve:

\begin{figure*}[t]
\epsscale{0.8}\plotone{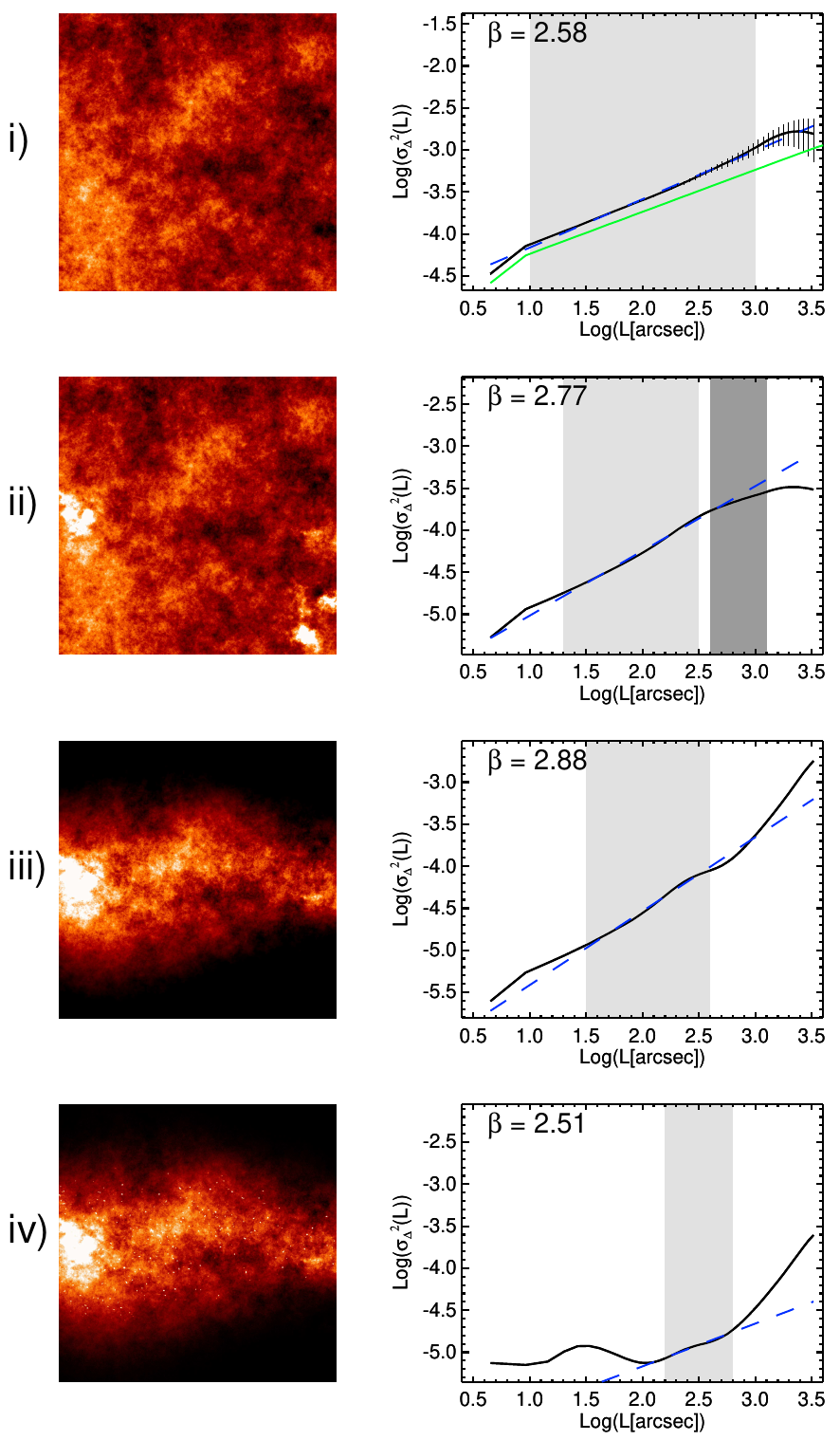} \caption{Simulation of a Hi-GAL image in
four main steps (left panels), and corresponding $\Delta$-variance plots (right
panels). Step~i): a $1800\times1800$ pixel fBm image is generated as a background.
Step~ii): the brightness of few limited regions is enhanced. Step~iii): the characteristic
shape of the Galactic plane is introduced through a Gaussian profile. Step iv):
500 Gaussian compact sources are spread across the map, and a low-level white noise
is finally added. The $\Delta$-variance curves are plotted in the right panels.The
$x$-axis variable is the decimal logarithm of the spatial lag $L$ (expressed in
arcseconds), and the range is the same for all panels. The range on the $y$-axis
is different for each panel instead, but its width has been kept the same in all
panels, to allow a visual comparison of the slopes. The $\Delta$-variance is represented
as a black solid line, while the linear fit over the inertial range (i.e. the range
where the curve has a power-law behavior, here highlighted with a grey area) is plotted
as a blue dashed line. In the top right panel, the $\Delta$-variance of the larger original
fBm set is also plotted as a green solid line. In the same panel, the error bars associated
to the $\Delta$-variance values are shown by way of example. \label{dvarsimul}}
\end{figure*}

\begin{itemize}
\item[i)] A $2700\times2700$ pixel fBm background has been generated with a
``typical'' power spectrum slope $\beta=2.5$ \citep[see, e.g.,][]{sch11}. Therefore,
to avoid dealing with a periodic image (see Section~\ref{fbm_sub}), this
has been truncated, extracting a sub-image of $1800\times1800$ pixels.
\item[ii)] To simulate the presence of very bright small regions, another
$2700\times2700$ pixel fBm set has been generated with a significantly higher
power spectrum slope ($\beta=3.4$) and a different phase distribution. The resulting
image has then been exponentiated in order to enhance the high-signal regions. Again, a
sub-image of $1800\times1800$ pixels has been extracted, making sure that the extracted
image still contains the maximum of the original image. Finally, this has been added
to the image obtained in i), resulting in regions of enhanced brightness.
\item[iii)] To reproduce the luminosity decrease off the Galactic plane,
a modulation through a Gaussian profile has been applied. To make this profile
more realistic, both the FWHM of the Gaussian and position of
the peak slowly float as a function of the longitude, following a Gaussian
distribution and a long-period sinusoid, respectively. A relevant decrease
of emission moving away from the plane is more pronounced in the Hi-GAL
observations of the inner Galaxy with respect to those considered in this work.
However, the goal of this test is to identify qualitatively the effect on
the $\Delta$-variance curve of peculiar structures , thus it is instrumental
to exacerbate these contributions.
\item[iv)] A population of 500 compact sources has been spread across the map,
generating random 2-dimensional Gaussians whose size and peak flux distributions
follow those found for the Hi-GAL field $\ell = 30^{\circ}$ \citep{eli10}. The probability
of displacing a source in a given position of the map has been weighted with the
intensity of the image in that position, to obtain a more realistic concentration
of compact sources in regions with bright diffuse emission. Before co-adding the
sources, the background image has been scaled by a given amount, such that its
dynamical range gets similar to that of the $\ell = 30^{\circ}$ background at $160~\mu$m.
Finally the image has been convolved with the PACS $160~\mu$m beam, and a
low-level white noise is summed over, using an additional fBm image with $\beta=0$.
\end{itemize}

In the right panels of Figure~\ref{dvarsimul}, the $\Delta$-variance of the
simulated images displayed on the left is plotted. All the images have been
normalized between 0 and 1, to prevent overflows in calculation. As a
consequences, the units of $\Delta$-variance are arbitrary. We note that the
$\Delta$-variance slope is not affected by this rescaling. In step i), the
extracted sub-image curve (black line) shows a slightly steeper slope than
that of the original set, plotted as reference (green line). While the latter
looks as a line as expected, the former exhibits a linear behavior only over a
limited range of scales (the grey area). The flattening of the curve at $L \gtrsim
2000\arcsec$ is just due to the truncation of the original fBm set. Moving to
step ii), a slope similar to the original one is still found on the limited range of
scales $400 \arcsec \lesssim L  \lesssim 1300\arcsec$ (darker grey area $\beta=2.45$),
while the steeper slope at $L \lesssim 400 \arcsec$ is caused by the co-addition of the
bright spots in the image. A further steepening is produced in step iii) by
modulating emission with a low spacial frequency profile. This is particularly
evident at the largest scales, where it compensates the flattening of the
$\Delta$-variance seen in the previous steps. Finally, the insertion of compact sources
in step iv) is responsible for the appearance of a bump for $10\arcsec \ll L \ll 100\arcsec$,
which is the typical size range of the injected sources.
This clearly corresponds, in light of the correspondence between
$\Delta$-variance and power spectrum, to the $\mathcal{P}_{\mathrm{cirrus}}(k)$
component of the power spectrum discussed by \citet{miv07} and \citet{mar10}.
Here the diffuse emission behavior can be recovered only for a limited
range of scales, since at the largest scales the effect due to
the Galactic plane shape is obviously still present. At the smallest scales,
instead, a flattening of the $\Delta$-variance curve is seen, due to white
noise  \citep[cf.][]{ben01}. In any case, no physical information can
be extracted at scales smaller than the instrumental beam.

Notice, however, that the significance of the effects described in this
section depends on the analyzed map. For example, for the third Galactic
quadrant maps we don't expect a strong influence of the Galactic plane
shape. Furthermore, the bump in $\Delta$-variance due to compact sources
is likely not as sharp as seen in Figure~\ref{dvarsimul}, because in the
real maps the transition between a compact source and the surrounding
cirrus emission is smoother than in our simulations. Larger clumps,
H\textsc{ii} regions and filaments present in the real maps are additional
intermediate structures between the two scale regimes of compact sources and the
diffuse emission, contributing to enlarge the bump toward larger spatial scales,
and to connect it more smoothly with the linear portion of the $\Delta$-variance.
The importance of the filaments in the scenario of star formation and their
ubiquity in the ISM have been highlighted by \emph{Herschel} observations
\citep[e.g.,][]{mol10b,arz11,rus13,sch13}, and they are certainly the main responsible
for the departure of the images from a fbm-like behavior at intermediate scales
between compact sources and cirrus.

\section{$\Delta$-variance of the third Galactic quadrant Hi-GAL fields}\label{dvarhigal_par}
\begin{figure*}[t!]
\epsscale{0.9}\plotone{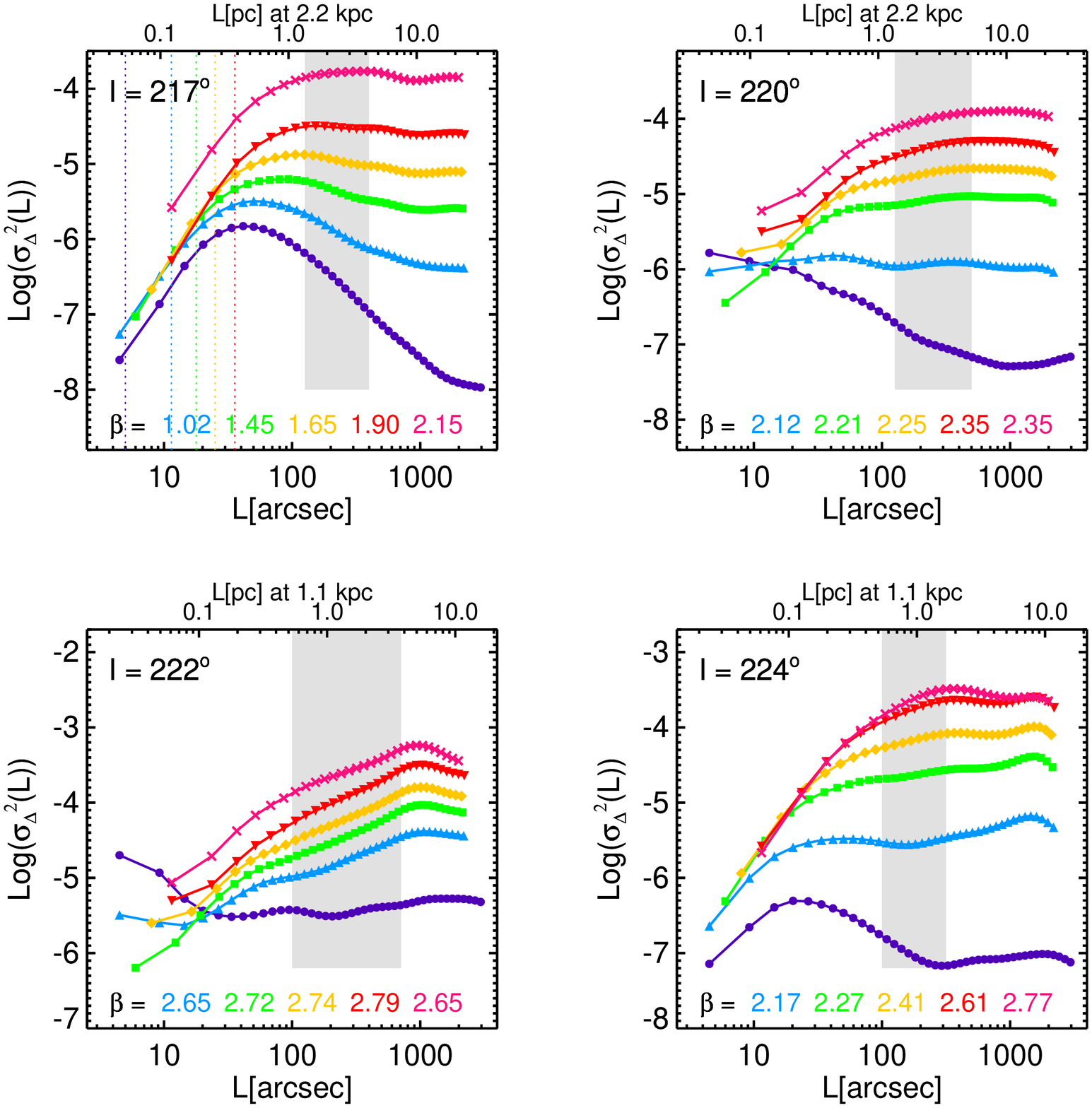}
\caption{$\Delta$-variance curves of the maps shown in Figure~\ref{higal2},
using the same tile naming and band-color encoding. Also in this case we
use the same $x$-axis range in all panels, and the same range width for the
$y$-axis, to allow a direct comparison of the slopes. The inertial range is
highlighted as a grey area. The corresponding linear slopes are transformed in
power spectrum slopes through Equation~\ref{dvar_beta} and reported on the
bottom of each panel (except for 70~$\mu$m). Finally, for reference
the spatial scales corresponding to the nominal beams at each band are
plotted as dotted lines in the top-left panel; scales below them are
meaningless.\label{dvarhigal}}
\end{figure*}

After normalizing, as described in Section~\ref{compact}, the Hi-GAL maps shown
in Figure~\ref{higal2}, the $\Delta$-variance vs spatial lag curves have
been computed. They are plotted in Figure~\ref{dvarhigal}
using the color-band coding introduced in Figure~\ref{higal2}.

\subsection{General results}\label{general_dvar}
Some general considerations can be drawn from the global trends exhibited
by these curves. The lack of a sufficient level of diffuse emission in the
$70~\mu$m maps influences the corresponding $\Delta$-variance spectrum,
which shows peculiar trends compared with other wavelengths. Therefore, the
curves at this wavelength are plotted only for completeness and the corresponding
slopes are not shown. Only as a general remark, we notice that in the tiles
$\ell$217 and $\ell$224, namely those with a significant emission of compact
emission at $70~\mu$m, a bump peaking at about 40\arcsec and 25\arcsec, respectively,
is present. At larger spatial scales, the $70~\mu$m curves show in all cases slopes
smaller those of the larger wavelengths, even negative in three cases out of four.
This behavior is expected in presence of low signal-to-noise ratio in the maps,
keeping in mind Equation~\ref{dvar_beta} and the white noise ($\beta$=0) borderline
case for the power spectrum slope.

For the remaining wavebands and for the column density maps, only one common
linear range has been identified for each tile. Noteworthy, scales shorter than
100$\arcsec$ have been not considered because of the possible contamination by
compact sources. The curvature of each line, defined as
$c = (\sigma_{\Delta}^2)''/\left\{1+[(\sigma_{\Delta}^2)']^2\right\}^{3/2}$
has been estimated, by allowing only ranges where reasonably low values of
$|c|$ are found\footnote[3]{In reality, only few $\Delta$-variance curves exhibiting
no linear behavior in the common interial range,
e.g. the one of $160~\mu$m of $\ell$224, have not been
considered in the evaluation of the linear range. However, despite these exceptions,
the constraint we apply on the linear range appears more robust than the qualitative
criteria adopted in the literature.}. This procedure keeps
the extremes of the range far from possible peaks of the curve (e.g. in $\ell$222),
where the most relevant departure from linearity is expected.

The estimated fitting ranges and the corresponding slopes of the power spectrum (obtained
from the $\Delta$-variance ones through Equation~\ref{dvar_beta}) are reported in
Figure~\ref{dvarhigal}, and summarized in Table~\ref{dvartab}, together with the
corresponding fractal dimensions of the maps (derived through Equation~\ref{deltafrac}).
\begin{deluxetable*}{lccccccccccccc}
\tabletypesize{scriptsize}
\tablecaption{Power spectrum exponent and fractal dimension of the investigated Hi-GAL maps \label{dvartab}}
\tablewidth{0pt}
\tablehead{ \colhead{Field} & Distance  & Fit range & \multicolumn{5}{c}{$\beta$} & & \multicolumn{5}{c}{$D$} \\
\cline{4-8} 
\cline{10-14} \\
\colhead{} & \colhead{[kpc]} &\colhead{[pc]} & \colhead{160 $\mu$m} & \colhead{250 $\mu$m} & \colhead{350 $\mu$m} & \colhead{500 $\mu$m} & \colhead{Col. dens.} & &
\colhead{160 $\mu$m} & \colhead{250 $\mu$m} & \colhead{350 $\mu$m} & \colhead{500 $\mu$m} & \colhead{Col. dens.}  }
\startdata
$\ell$217 & 2.2 & 1.3--4.2 & 1.02 & 1.45 & 1.65 & 1.90 & 2.15 &  & 3.49 & 3.27 & 3.17 & 3.05 & 2.93 \\
$\ell$220 & 2.2 & 1.3--5.3 & 2.12 & 2.21 & 2.25 & 2.35 & 2.35 &  & 2.94 & 2.89 & 2.87 & 2.83 & 2.83 \\
$\ell$222 & 1.1 & 0.5--3.8 & 2.65 & 2.72 & 2.74 & 2.79 & 2.65 &  & 2.67 & 2.64 & 2.63 & 2.61 & 2.68 \\
$\ell$224 & 1.1 & 0.5--1.7 & 2.17 & 2.27 & 2.41 & 2.61 & 2.77 &  & 2.91 & 2.86 & 2.79 & 2.70 & 2.61 \\
\enddata
\end{deluxetable*}

As a general remark on some evident trends found in all the four tiles, we notice 
that, from the qualitative point of view, for each tile the three SPIRE
 bands have quite similar spectral behaviors, whereas
the general shape of the PACS 160~$\mu$m curves appear relatively different.
More importantly, a systematic increase of the slope with the wavelength is found from 
160 to 500~$\mu$m.This suggests that not only the emission morphology changes when observed 
at different wavelengths, but also its statistical properties are found to be different. In
particular, at 160~$\mu$m the contribution of warmer Very Small Grains can be still relevant 
\citep{com10} and seems to be responsible for a more uniform distribution of the power of the image 
through the different spatial scales, resulting in a shallower $\beta$. In other words, the warmer 
dust turns out to be more diffuse and spread around than the cold dust, which is expected to be 
preferentially concentrated in denser environments like filaments \citep[e.g.,][]{pad06,cam07},
producing a global smoothing of the observed emission features. 

This effect can represent
a possible explanation for the systematic discrepancy, found by \citet{sch11}, between
the $\Delta$-variance slope of $A_V$ and $^{13}$CO maps (shallower and steeper, respectively)
of the same areas of the sky. Most likely, these two tracers do not describe exactly the
same components of the ISM in the same manner \citep[e.g. ][]{goo09}. Moreover, the ISM
is optically thinner at the SPIRE wavelengths than at 160~$\mu$m, so at long wavelengths
one expects a more enhanced contrast between emission from high-density and low-density
regions, resulting in a possible steepening of the power spectrum (and, equivalently, of
the $\Delta$-variance). However this reasoning could be too simplistic, because moving
towards SPIRE wavelengths a number of cold small scale filaments can manifest themselves
thus contributing to the power spectrum and counter-balancing the effect described above.

Another consideration concerns the $\Delta$-variance of the column density
maps: although the maps look quite similar to the SPIRE ones, the power
spectrum behavior is generally found to be slightly different from
those. Anyway, all the slopes we found lie
in the typical range of values found for clouds studied in previous
works \citep{ben01,sch11,row11,rus13}.

\subsection{Results of individual maps}\label{maps_dvar}
Going into detail of single tiles, we start from the westernmost field,
$\ell$217, associated with distance component~I (see Figure~\ref{higal2}).
As in the case of $\ell$224 discussed in the following, the abundance of
compact sources and filamentd in this field produces a visible bump at
$L \lesssim 100\arcsec$ (see Section \ref{compact}) at all wavelengths,
whose peak and upper endpoint shift
towards larger scales with increasing wavelength. It is generally followed
by a descending trend which stops around 1000\arcsec, corresponding to a
physical scale of $\sim 11$~pc at a distance of 2.2~kpc. The presence of the large filament
associable with Sh~2-287 \citep{sha59} in the northern part of the map is probably
the responsible of this slope change, and of the departure of the maps from
self-similarity at the smaller scales.

The $\ell$220 tile exhibits a more extended range of linearity. In fact this tile is,
in the data set we consider, the poorest of bright features, so that the cirrus component
can be really probed. The inertial range (1.3-5.3 pc) partially overlap those found
by \citet{sch11} for some low-mass star forming clouds. A good correspondence is
also found with the \emph{Herschel}-based analysis of the NGC 6334 star forming region
($d=1750 pc$) of \citet{rus13}, where three out of four separate sub-regions exhibit
inertial ranges similar that of $\ell$220. At longer scales, around 15~pc, the
$\Delta$-variance curves flatten, which does not
seem to correspond to any visible structure in the maps, such as filaments,
ridges or bubbles (and corresponding cavities). In this case it is likely that
the upper limit of the self-similarity range is really correlated with the
injection scale of turbulence.

In tile $\ell$222 the steepest $\beta$ slopes are found, over an inertial
range of 0.5-2.7~pc. As in the case of $\ell$220, the small number of bright
features in the maps translates into a wide range of linearity of the $\Delta$-variance
and into a weak compact source bump.
A peak is present at $\sim 1000\arcsec$ ($5.4$~pc at $d=1.1$~kpc), corresponding
approximately to one sixth of the map size. Although we do not have sufficient
information to claim that this feature is associated with the bubble located on the
East side of the tile (see Figure~\ref{higal2}),
we believe that this structure is responsible for the high values of $\beta$ we find:
indeed it generates a certain degree of segregation between relatively empty and
bright regions (cf. for example panels $b$ and $d$ of Figure~\ref{fbms}), hence
a transfer of power towards large scales.

Finally for $\ell$224, similarly to $\ell$217, we find that the contribution of
compact sources and filaments introduces features in the $\Delta$-variance curves
which make it difficult to identify a possible inertial range. The one we find
between 0.5 and 1.7 pc (neglecting the 160~$\mu$m curve) is compatible with that
of $\ell$222. The slopes are generally steeper than those of the component II tiles,
but shallower than those of $\ell$222.

This comparison suggests that the region of the plane covered by the eastern
tiles ($\ell$222 and $\ell$224), which is quite coherent from the kinematical
point of view being associated with distance component~I, show some common
global statistical properties, different from those of the western tiles
($\ell$217 and $\ell$220, component~II), generally characterized by
shallower slopes. Furthermore, within the two distance components, the tiles
containing bright features ($\ell$217 and $\ell$224, respectively) show slopes
shallower than those of the corresponding low-emission tiles ($\ell$220 and
$\ell$222, respectively).

Anyway, the variety of power-spectrum slopes we find in different tiles
reinforces the scenario of the non-universality of the ISM fractal properties.
From the morphological point of view, this corresponds to a different distribution
of the power of the image on a spatial frequency range. From the point of view of
the underlying physics, different power spectrum slopes are related to different
conditions of the compressible turbulence which is widely considered the most
realistic situation in the ISM, especially in presence of star formation
\citep{hen84,padn02}. Unlike the ``rigid'' results of the \citet{kol41}
incompressible turbulence (see Section~\ref{fbm_par}), the compressible one
is able to produce, in the ISM, a variety of morphologies and hence of power
spectrum profiles \citep{fed09}.

%Notice that
%in this tile a better agreement between the slope at 160~$\mu$m and those at the
%longer wavelengths is found.

\subsection{The fractal dimension of the images}\label{fracdim}
The fractal dimension $D$ is another important observable, useful to characterize 
the ISM morphology. As mentioned in Section~\ref{intro}, several computational approaches 
have been adopted to derive it. Here we use Equation~\ref{deltafrac}, assuming that the 
analyzed maps have a fBm-like behavior in the recognized inertial ranges.
This makes the descriptions based on $\beta$ and 
$D$ completely equivalent, however to speak in terms of fractal dimension here allows us to 
make further comparisons with observational and theoretical results present in literature.
A check on possible alternative methods for calculating the fractal dimension is 
provided in Appendix~\ref{boxcount}.
 
The linear relation between the power spectrum slope $\beta$ and the fractal dimension
$D$ contained in Equation~\ref{deltafrac} clearly expresses the intuitive concept
that a ``smoother'' texture (high $\beta$, see Figure~\ref{fbms}) must correspond
to a lower degree of fractality (i.e. low $D$). The factor $\frac{1}{2}$ appearing
in the equation can half the perception of the variation of $D$, which in fact is
expected to vary only between 2 and 3. Although in the literature a variation of $0.1-0.2$
between two values of $D$ is presented as negligible, it actually corresponds to significant
structural differences of the maps. In this respect, the fractal dimensions reported
in Table~\ref{dvartab} reassert 
again i) the dicrease of $D$ at increasing
observed wavelength (already discussed as increase of $beta$ in Section \ref{general_dvar}),
and ii) the significant differences between the structure observed in the western and in 
the eastern tiles.

In the three tiles showing fractal behavior ($\ell$220, $\ell$222 and $\ell$224),
all the values we derived range from $D=2.61$ (500~$\mu$m of $\ell$222 column density
of $\ell$224) to $D=2.94$ (160~$\mu$m of $\ell$220), and most of them are compatible
with the typical range of variability found through statistical techniques
\citep[][and references therein]{row11,sch11}, but also in most part with the
$2.6 \lesssim D \lesssim 2.8$ found by \citet{san07,san09} through the perimeter-area
relation.

Instead, the values we obtained are significantly larger than the average value
found on IRAS 100~$\mu$m maps by \citet{miv07} ($\beta=2.9$, which is $D=2.55$
in the fBm approximation), in particular those found at 160~$\mu$m, which is
the closest band to the IRAS 100~$\mu$m one.

In any case, the values we find are far away from the first estimates of the fractal
dimension of interstellar clouds \citep[$D=2.3$,][]{fal91,elm96,elm97b}, which
initially described a quite constant and universal behavior of the ISM structure,
and were found also to be in good agreement with the predictions of the Kolmogorov
incompressible turbulence \citep[$\beta=3.6$ for the density field, then $D=2.2$,][]{fal94}.
Instead, the values found here, as well as the others obtained through the
$\Delta$-variance \citep[][and references therein]{sch11}, are more compatible with
the power spectra slopes of compressible turbulence, \citep[e.g.][]{fed09}, and in
particular for the case of solenoidal driving of turbulence ($\beta=2.89$, $D=2.55$).

We can also compare with the theoretical predictions of \citet{gaz10} on
the power spectrum of both 3-dimensional and column density in turbulent
thermally bistable flows\footnote[4]{A gas in which temperatures above and
below a given instability range can coexist in thermal pressure equilibrium is
called bistable \citep[see][and references therein]{gaz05}. The atomic ISM is
generally believed to be bistable.}. Their slope (calculated between 7 and 25~pc,
i.e. a range of scales larger than ours) gets shallower as the Mach number
increases: from $\beta=2.64$ at $\mathcal{M}=0.2$ to $\beta=2.11$ at
$\mathcal{M}=4.0$ \citep[a similar trend can be clearly found also in ][]{kri06,kow07}.
Although these simulations essentially reproduce the H\textsc{i} emission and
despite the different investigated spatial ranges, the behavior we find suggests
that the two eastern fields are characterized by smaller Mach numbers. In turn,
the star formation efficiency of a cloud is recognized to inversely depend on
the Mach number, whose contribution, however, can be found to be concomitant and
then surpassed by that of other physical factors \citep[e.g.,][]{ros10}.

\begin{figure*}[th!]
\epsscale{1.2}\plotone{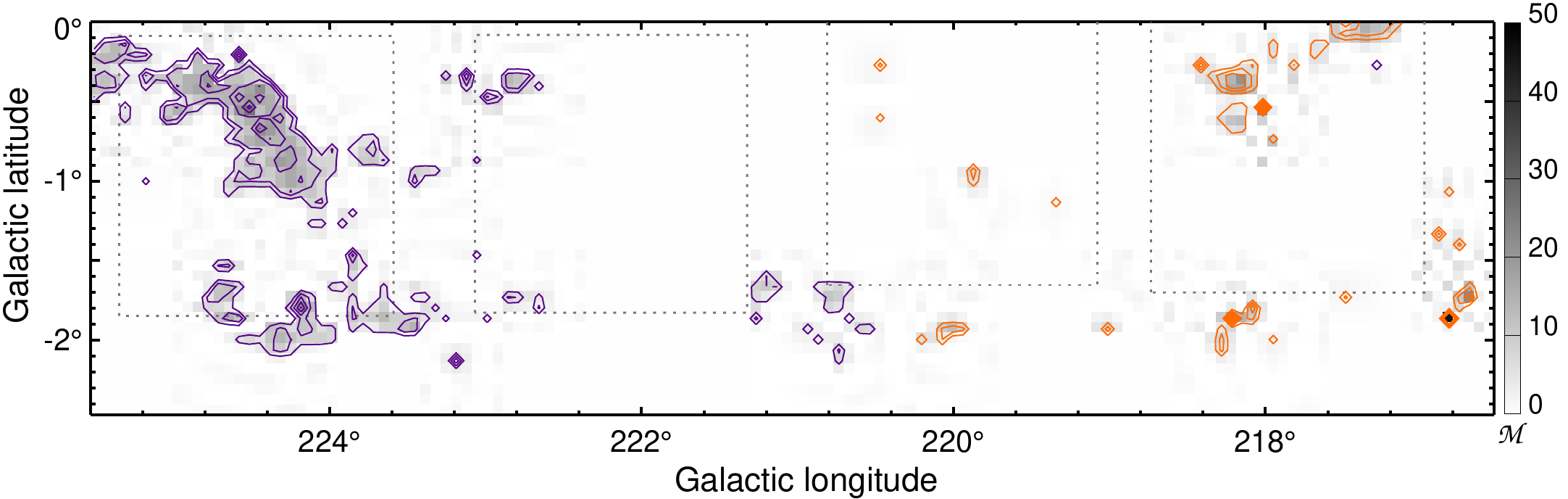}
\caption{Greyscale map of the Mach number for distance components~I and~II (identified with purple
and orange contours, respectively) of \citetalias{eli13}, calculated across the entire
NANTEN available data cube as described in the text. Contours range from 5 to 30 in steps
of~5. The grey dotted squares enclose the four areas investigated
in this paper.\label{machfig}}
\end{figure*}

To investigate this aspect in our data, we derived the 3-D turbulent Mach
numbers from the NANTEN data cube for the velocity components~I and~II,
as the ratio between the local values of the deconvolved CO(1-0) line width
$\left\langle\sigma_{CO}\right\rangle$ and the sound speed $c_s$:
\begin{equation}\label{mach}
  \mathcal{M}=\sqrt{3}\frac{\left\langle\sigma_{CO}\right\rangle}{c_s} \; .
\end{equation}
Here $c_s=(kT/\mu)^{1/2}=0.188(T/10 \mathrm{K})^{1/2}$~km~s$^{-1}$, where $T$ is taken
from the dust temperature maps obtained in \citetalias{eli13} simultaneously with the column
density maps. The amount of CO spectra suitable for this analysis is limited by few constraints.
First, only lines with a reasonable SNR are considered \citepalias[see details in][]{eli13};
second, the line full width at half maximum must be larger than the spectral resolution element
($\Delta v=1$~km~s$^{-1}$) to allow deconvolution; third, the location CO spectra must lie
within one of the four fields we investigate in this paper. These are shown in Figure~\ref{machfig}
(left panel), overlaid to the distribution of the Mach number across the entire NANTEN data set.

%\begin{figure*}[th!]
%\epsscale{1.2}\plotone{mach.eps}
%\caption{Left: greyscale map of the Mach number for components I and II (identified with purple and
%orange contours, respectively) of \citetalias{eli13},
%calculated across the entire NANTEN available data cube as described in the text. Contours range from
%5 to 30 in steps of 5. The four grey dotted squares represent the boundaries of the four
%areas investigated in this paper. Right: Mach number distribution of the two velocity components,
%based on the CO pointings lying inside the four squares of the left panel. Color convention
%for the components I and II is the same used in the left panel.\label{machfig}}
%\end{figure*}

We note that we can derive meaningful statistical information only for the brightest
regions, while the observations are not sensitive enough to trace the cirrus component,
which is the one showing the self-similar behavior that justifies such a test.
%Apart from the different number of involved lines, the two Mach number distributions we
%obtain (Figure~\ref{machfig}, right) appear statistically similar, as testified also by
%their averages: $\left<\mathcal{M}_I\right>\pm \sigma_{\mathcal{M}_I}=11.8\pm 3.6$ and
%$\left<\mathcal{M}_{II}\right>\pm \sigma_{\mathcal{M}_{II}}=11.1 \pm 4.5$, respectively.
For this reason, we stress that observations with better sensitivity and spectral resolution
are required for confirming this result.

We recall that in \citetalias{eli13} the distance components~I and~II have been
found to differ in star formation efficiency by a factor 2 (0.008 against 0.004,
respectively). Interestingly, a direct correspondence between a larger fractal
dimension and a higher star formation efficiency is found in this case. Unfortunately,
the inconclusive argument of the Mach number can not support the aforementioned
theoretical picture, so that further checks are required and will be feasible when
star formation efficiencies will be available for other portions of Hi-GAL.

From this discussion it clearly emerges that the estimation of the fractal dimension
of far-infrared dust emission from interstellar clouds (or portions of them) constitutes
a further constraint to be put on theoretical models of cloud structure. In
this sense, the vast amount of data provided by the multi-wavelength Hi-GAL
survey represents a real minefield which can be searched for such observables.

\subsection{Cloud structure and clump mass function}
The quantitative indications obtained in the previous sections allow us to probe a
relation derived by \citet{stu98}, linking the power spectrum of the ISM to the mass
function of the clumps (CMF hereinafter) which originate from such a structure. Indeed,
as mentioned in Section~\ref{intro}, these authors showed that the fractal description
of the ISM and the approach based on a hierarchical decomposition \citep[e.g.,][]{hou92}
are compatible. \citet{stu98} demonstrated that, for a fBm-like cloud characterized by
a power-law power spectrum with slope $\beta$, and assuming a clump mass spectrum in the
cloud $dN/dM\propto M^{-\alpha}$ (in its high-mass end) and a relation between the clump
mass and size as $M\propto r^{\gamma}$ , then
\begin{equation}\label{clumpbeta}
\beta=(3-\alpha)\gamma \;.
\end{equation}

This theoretical relation contains the intuitive concept that, once $\gamma$ is fixed
(where the cases $\gamma=3$ and $\gamma=2$ indicate clumps with constant volume and column
density, respectively), a steeper power spectrum corresponds to a power transfer towards
larger scales, hence to the presence of bigger clumps, which makes the CMF shallower.

So far Equation \ref{clumpbeta} has been tested on very few data sets, due to the
lack of observational constraints, namely the simultaneous knowledge of the power
spectrum and of the CMF of a cloud. \citet{stu98} combined the ``average''
slopes $\beta$ and $\alpha$ found for a sample of molecular clouds ($\beta=2.8$
and $\alpha=1.6-1.8$, respectively), deriving $\gamma=2-2.33$. More recently,
\citet{sha11} explored this relation analyzing 3-dimensional fBm synthetic
clouds, starting from the assumption that molecular clouds have on average
$\beta=2.8$ (different, however, from the shallower slopes we find in $\ell217$
and $\ell220$).

\begin{figure}[t!]
\epsscale{1.}\plotone{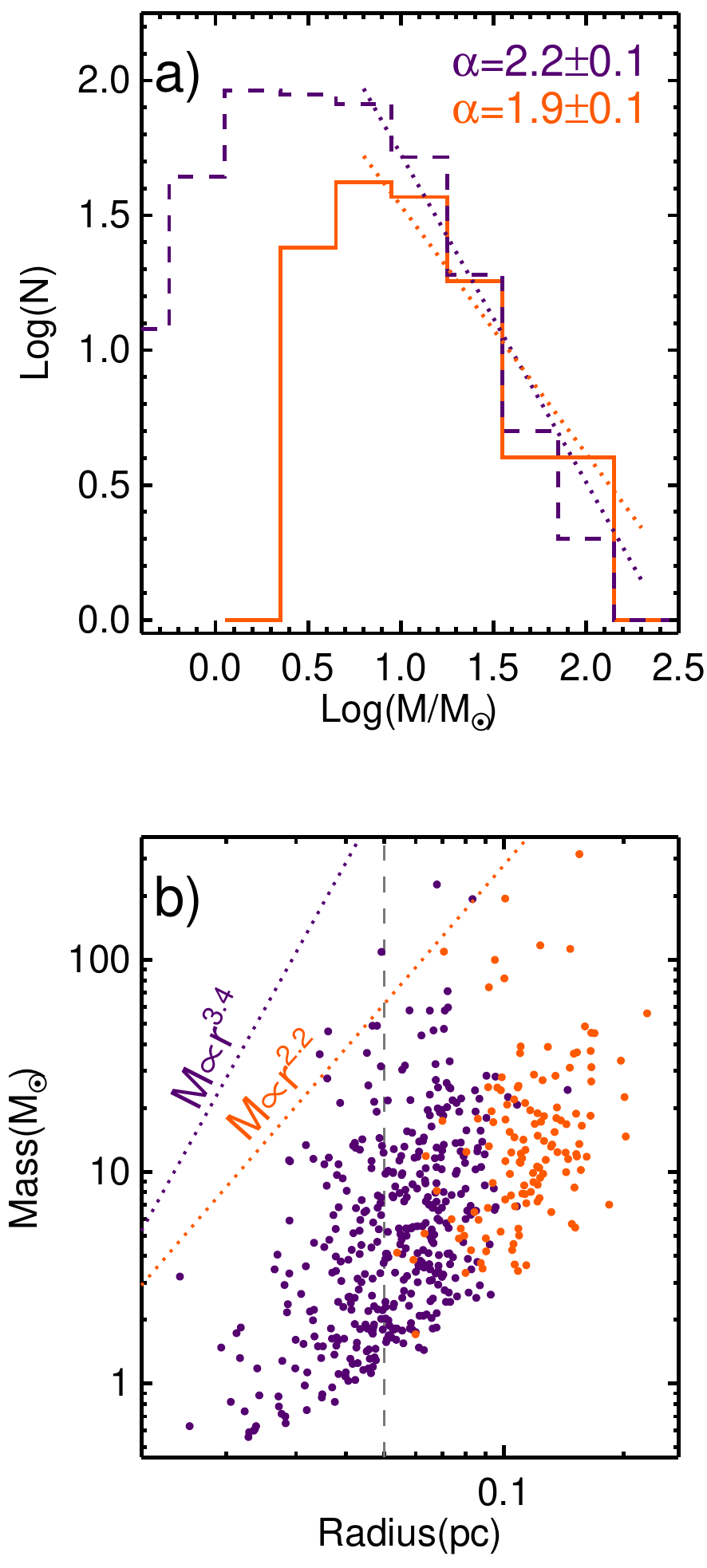}
\caption{Panel $a$: CMFs (solid histograms) of the pre-stellar clumps associated
to the velocity components~I (purple) and II (orange), respectively. Being the CMF
in logarithmic form, the slope of the linear fit (dotted lines) corresponds to
$\alpha-1$, with $\alpha$ defined as in the text.
Panel $b$: mass vs radius plot for the sources associated to components I and II
(same colors as in panel $a$). The dotted lines represent the linear trends expected
from Equation~\ref{clumpbeta} (not to be confused with a fit to data). The grey
dashed vertical line expresses the ideal transition between cores and clumps,
according to the classification of \citet{ber07}.
\label{cmf}}
\end{figure}

Here we have the possibility to check this relation using the estimates of $\beta$ for
the investigated fields, and exploiting the clump masses derived in \citetalias{eli13}
to build a statistically significant CMF by selecting the sources as follows. First, it
makes sense to consider only pre-stellar cores, namely starless sources gravitationally
bound \citep[i.e. those exceeding their corresponding Bonnor-Ebert mass, see e.g.][]{gia12}.
Second, the considered source samples should be coherent from the spatial point of
view, namely all the sources should belong to a physically connected region, following
the steps described in \citetalias{eli13} for 398 compact sources associated with the
closest velocity component (I), which is the component dominating the $\ell222$ and
$\ell224$ fields. Here we consider also component~II for comparison, identifying
131 sources suitable for this analysis. The amounts of
sources associated with components~III and~IV, instead, are not statistically relevant
to extend the analysis. The slope of the CMF for component~I has already been estimated
in the \citetalias{eli13} as $\alpha_I=2.2\pm0.1$, using an algorithm independent of the
histogram binning \citep[][and references therein]{olm13} which also allows us to determine
the lower limit of the validity range of the power law. Similarly, for component~II
here we obtain $\alpha_{II}=1.9\pm0.1$ (Figure~\ref{cmf}, panel $a$).

For the power spectrum slopes, we decided to use the values obtained for the column
density maps of the two tiles characterized by low contamination of compact sources,
namely $\ell$220 for component I and $\ell$222 for component II, respectively.
Thus we assume $\beta_I=2.65$ and $\beta_{II}=2.35$ for components~I and~II,
with error bars of 0.1, and we get $\gamma_I=3.4\pm 0.2$ and $\gamma_{II}=2.2 \pm 0.1$,
respectively.

A comparison between these results and the observational mass vs radius
relation can be performed using the same sources identified for building
the CMFs of components~I and II (Figure~\ref{cmf}, $b$). However, the
distributions are affected by a significant dispersion and lack of a clear power-law
trend, as noted in previous analysis of this kind \citep[e.g.,][]{and10,gia12}, so
no clear-cut conclusion can be reached. Indeed, the lack of a definite scaling
relation is indicative of a departure from pure self-similarity. Moreover, despite
such clumps have been selected in order to build samples as homogeneous as possible,
residual differences of physical and evolutionary conditions can be encountered.
This might increase the scattering in the mass vs size plot, making it rather difficult
to confirm the theoretically expected value for $\gamma$.

It is interesting, instead, to discuss the two very different values of $\gamma$ we find,
especially in the context of the applicability of Equation~\ref{clumpbeta} to various
cases. The large discrepancy between these values depends on the different average $\beta$
found for the two components, but mostly on the different values of $\alpha$ appearing
at the denominator of Equation~\ref{clumpbeta}. The discrepancy between the CMF slopes of
the two components can be justified, in turn, with the relatively different intrinsic nature
of the observed sources. \citetalias{eli13} discussed the fact that the detected compact
sources, characterized by the same range of angular sizes (from one to few instrumental point
spread functions) correspond however to a variety of physical sizes, depending on the source
distance. In this case the typical factor $\sim 2$ between distances of component~I and~II
sources is responsible of the clear segregation in size between the two populations, visible
in Figure~\ref{cmf}, panel~$b$, such that the former are clumps, while the latter are a mixture
of clumps and cores, according to the typical definition of these categories \citep[e.g.,][]{ber07}.
It is widely acknowledged that the mass function slope of the cores is similar to that of the
stellar initial mass function \citep[e.g.,][]{and10} and steeper than that of larger clumps
\citep[e.g., $\alpha=1.6-1.8$,][]{kra98}, as a consequence of the underlying
fragmentation process. Samples characterized by contamination generally present
intermediate slopes \citep[e.g.,][]{gia12}. In this scenario, the CMF slope of the
component~II looks more similar to the case of the Polaris flare CO clumps
\citep[$\alpha=1.8$,][]{hei98} used by \citet{stu98} to derive a value of $\gamma=2.3$.
The component~I source sample, on the contrary, consists of sources intrinsically smaller
(as better resolved), closer to spatial scales where gravity dominates the morphology
of the ISM, definitely stopping its scale-free behavior. As a consequence,
in this case the applicability of Equation~\ref{clumpbeta} turns out to be dubious.

These indications must be confirmed in future by extending this kind of analysis
to a larger variety of cases, offered by further Hi-GAL fields characterized by
favorable observing conditions, typically found in the outer Galaxy.

\section{Summary}\label{summary}

We have analyzed the first four fields of the outer Galaxy
($217^{\circ} \lesssim \ell \lesssim 225^{\circ}$, $-2^{\circ}\lesssim b \lesssim 0^{\circ}$)
observed by \emph{Herschel} as part of the Hi-GAL survey, to give a
quantitative description of the structure of the far infrared diffuse
emission. We exploited the power spectrum slope as descriptor, and the
$\Delta$-variance algorithm as a tool to derive it in a robust way.
The low degree of confusion along the line of sight,
revealed by CO line observations, and both the resolution and the
large size of the \emph{Herschel} maps represent an ideal case to study
the structure of the diffuse ISM on the Galactic plane. In this
respect, the results of our analysis have a double value, because on
one hand we characterize the morphology of the ISM in this portion
of the plane, on the other hand we provide a set of general
prescriptions, considerations and caveats for a possible
extension of this kind of analysis to other \emph{Herschel} maps.
It is difficult in some cases to separate global from local results.
Accordingly, in the following we try to summarize the conclusions going
from the more general ones to those which have a more specific value:
\begin{itemize}
  \item[-] The presence of compact sources in the maps affects the
  $\Delta$-variance curves in terms of a bump up to scales of $\sim 100\arcsec$.
  Other relevant effects are produced by the interruption of the self-similarity
  due to the Galactic plane latitude emission profile, and, on smaller spatial
  scales, to the presence of bright filaments and star forming regions. In our
  sample, however, these effects (especially the former) are less relevant
  than, for example, in the Hi-GAL fields of the inner Galaxy.
  \item[-] The maps obtained in four PACS/SPIRE wavebands (160-500 $\mu$m)
  for the investigated fields, together with the column density maps derived
  from them, show common features of the $\Delta$-variance (or, equivalently,
  of the power spectrum), i.e. spatial scales corresponding to peaks and turn
  over points. However, the slope of the power spectrum can remarkably change
  from one band to another, generally increasing from 160~$\mu$m to 500~$\mu$m,
  probably due to a different spatial displacement of small against large
  grain dust components.
  \item[-] The obtained power spectrum slopes strongly vary from one tile to
  another, but remain in the typical range of slopes $2 \lesssim \beta \lesssim 3$
  calculated for different phases of the ISM by \citet{sch11}. The power spectrum,
  and the corresponding fractal dimension, are then far from being constant and
  universal as initially suggested by the analysis of the boundaries of the
  interstellar clouds \citep{fal91}. In two of the four investigated fields,
  the presence of regions of intense emission is responsible of a peak in the
  power spectrum, followed by a linear trend with slope shallower than the case
  of the other two tiles, dominated by the cirrus.
  \item[-] None of these slopes is consistent with the Kolmogorov's
  incompressible turbulence case. We find that the model of supersonic isothermal
  turbulence with solenoidal forcing of \citet{fed09} seems more realistic,
  as already found by \citet{sch11} using different tracers.
  \item[-] The power spectrum slopes of the two eastern fields, $\ell224$
  and $\ell222$ are steeper than those of the western fields, $\ell220$ and
  $\ell217$. Each of these two pairs is dominated by the emission of a
  different distance component ($d=1.1$~kpc and 2.2~kpc, respectively).
  Morphological differences of the density field suggest a higher degree
  of turbulence in the second component, also suggested by a lower star
  formation efficiency (yet not confirmed by the Mach number analysis).
  This represents an interesting evidence of the connection between fractality
  and star formation efficiency predicted by the theory.
  \item[-] We tested the linear relation, predicted for fBm-like clouds, between
  the mass function of the over dense structures (clumps) and the cloud power
  spectrum slope, using as a probe the exponent of the clump mass vs radius
  power law relation. However the mass vs radius distribution we obtain is
  affected by significant dispersion and can not be used to draw strong conclusions.
  Also, this relation seems to work better on the clumps at 2.2~kpc, while
  its predictions at 1.1~kpc are unrealistic, probably because the sample of
  clumps associated to this component is contaminated by cores.

%%%  \item[-] Analyzing in more detail the $\ell217$ tile at two complementary
%%%  wavelengths, 160~$\mu$m and 500~$\mu$m, different power spectrum behaviors
%%%  emerge in different positions of the region, with no apparent correlation
%%%  between map intensity and power spectrum slope. Instead, for the 160~$\mu$m
%%%  band, more sensitive to the temperature, the change of slope occurring at
%%%  $\sim 100\arcsec$ (1~pc at a distance of 1100~pc) increases with map emission
%%%  and also with the average temperature in the range $14.0~K < T < 14.7~K$. This
%%%  evidence can be tentatively explained invoking the action of turbulent heating
%%%  due to dissipation occurring at the characteristic scale of the slope turnover.
\end{itemize}

\acknowledgments
Data processing and map production have been possible thanks to
generous support from the Italian Space Agency via contract
I/038/080/0. Data presented in this paper were also analyzed
using the \emph{Herschel} interactive processing environment (HIPE),
a joint development by the \emph{Herschel} Science Ground Segment
Consortium, consisting of ESA, the NASA \emph{Herschel} Science
Center, and the HIFI, PACS, and SPIRE consortia.

During the preparation of this paper, D. E. lost his father, Mr. Elio
Elia. D. E. is grateful for the example he set and his inborn and unconditional
generosity. Nothing in D. E.'s career, including this paper itself,
would have been possible without his father's steady support in any
aspects of his life. This paper is dedicated to his memory.


%% in parentheses, after the keyword, but they will not be verified.

{\it Facilities:} \facility{Herschel (PACS, SPIRE)}.



\appendix
\section{Box counting fractal dimension}\label{boxcount}
In this paper we followed the fBm approach to derive the fractal dimension 
$D$ of the analyzed images from the analysis of their power spectrum, through 
Equation~\ref{deltafrac}. Several other approaches might be used, however, to derive 
$D$ for grayscale images. These can be seen as surfaces defined on a 2-dimensional
plane and embedded in a 3-dimensional space ("mountain surfaces"). A fractal image,
in particular, is a complex surface whose dimension is larger than 2 (as the Euclidean 
geometry would expect).

Methods aimed at estimating the fractal dimensions of the image intensity 
contour levels, as the \emph{perimeter-ruler} and the \emph{area-perimeter} relation, 
are based on the distorted assumption that the image fractal dimension can be derived 
from the contour dimension simply by adding 1. This appears immediately wrong in 
the case of highly anisotropic fractals, but also in more isotropic cases, as the fBm 
sets (see Equation~\ref{deltadiff}) and simulated IS clouds \citep{san05}.
 
Instead, considering the fractal set itself, it is possible to use approaches based on
directly sampling the set with grids having different spacings. For example, the \textit{box counting} 
approach is one of the most intuitive ways to derive the fractal dimension, and to
understand its sense as well. Following \citet{man83}, a 3-dimensional cube containing the fractal
surface $A(x,y)$ is partitioned into grids of cubes of variable size 
$L$. The number of volume elements occupied by the fractal set is a function of the element size: 
$N(L)\propto L^{-D_{B}}$, where $D_B$ designates the box-counting fractal dimension, which can be 
obtained from the least square linear fit of the logarithms. 
However, whereas the estimate of $N(L)$ is rather trivial for binary fractal sets, it 
requires additional care in the case of gray-scale images as ours. One possible approach consists in
partitioning the $(x,y)$ plane in a grid of squares (boxes) of size $L$, and in estimating the difference 
between the maximum and the minimum values (normalized to integer multiples of $L$) achieved by the 
image in each box \citep{sar94}. This is needed to estimate the \emph{volume} spanned by the image in such 
box. At each explored value of $L$, the average $\mu_L$ of the volume over all boxes is computed.
Again, the so-called \emph{mass fractal dimension} can be derived from the power-law scaling of
$\mu_L$:
\begin{equation}\label{boxfrac}
  \mu_L \propto L^{D_M}\,,
\end{equation}
\citep{man83}.

\begin{figure*}[h!]
\begin{minipage}{8.0cm}\begin{center}\plotone{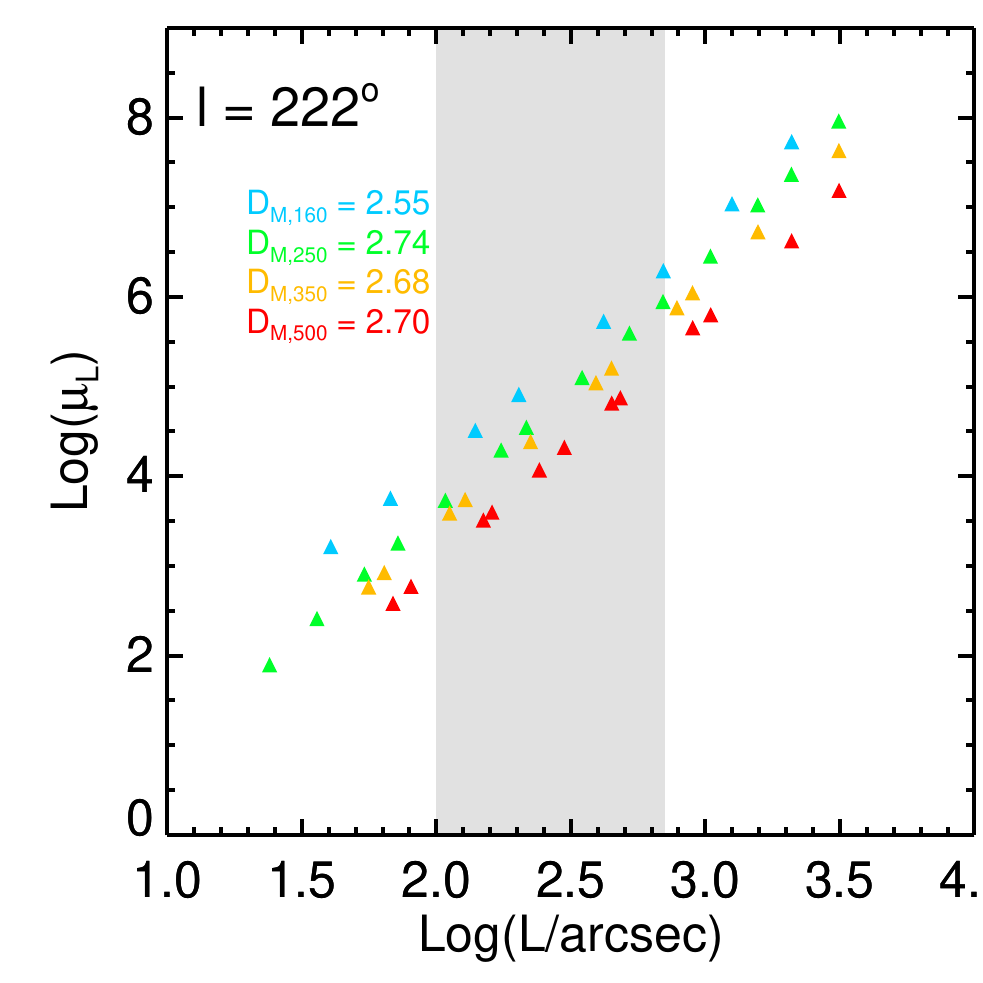}\end{center}\end{minipage}
\begin{minipage}{8.0cm}\begin{center}\plotone{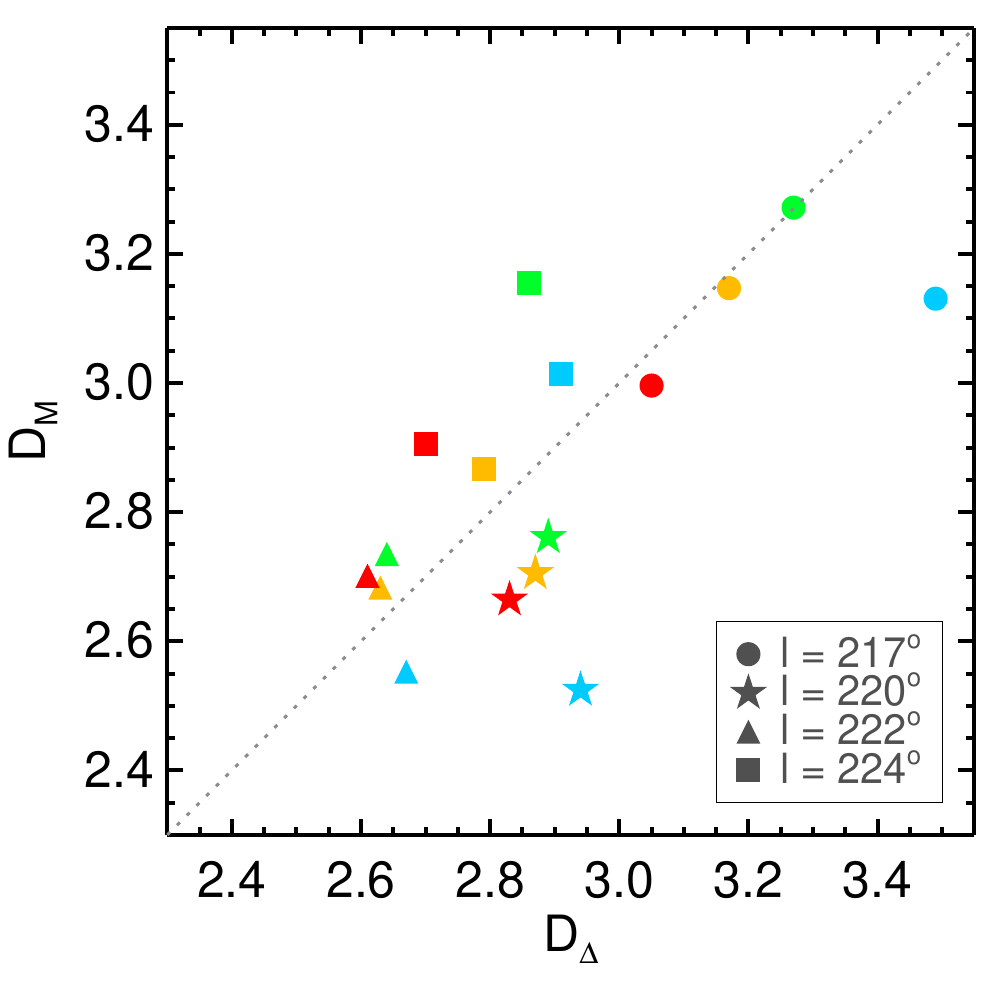}\end{center}\end{minipage}

\caption{\textit{Left}: average \textit{volume} (see text) occupied by the maps of $\l222$,
plotted as a function of the investigated box size (reported to units of arcseconds to be comparable 
with the lag~$L$ used throughout the article). Tile naming and band-color encoding are the same of
Figure~\ref{higal2}. Fractal dimensions correspond to the linear fit slopes of these series, estimated
over the same scale range of Figure~\ref{dvarhigal}, bottom-left panel. They are reported in different
colors, according to wavelength. \textit{Right}: fractal dimensions of the maps analyzed in 
this paper (tiles from $\ell$217 to $\ell$224, wavelengths from 160 to 500 $\mu$m), derived
through $\Delta$-variance ($x$ axis, from Table~\ref{dvartab}, here indicated with $D_{\Delta}$ 
to avoid confusion) and box counting techniques ($y$ axis, this Appendix). The correspondence 
between tiles and symbol types is reported in the box at the bottom right corner. The grey dotted 
line represents the bisector of the plane. \label{fdims}}
\end{figure*}

For each tile and at each wavelength between 160 and 500 $\mu$m, we evaluated the linear fit in 
the same spatial scale range reported in Table~\ref{dvartab}. In Figure~\ref{fdims},
left panel, an example of the $D_M$ calculation is shown for the $\ell$222 tile at four different 
wavelengths.

The comparison of the box-counting vs. the $\Delta$-variance-based fractal dimension of all
the Hi-GAL maps analyzed in this article is shown in Figure~\ref{fdims}, right panel. In most cases,
a good agreement is found between the two methods at the SPIRE wavelengths, whereas strong discrepancies
are found for PACS 160~$\mu$m, in at least three tiles out of four. It is possible that, being the 160~$\mu$m 
band the one exhibiting the most evident departures from a linear (i.e. fractal) behavior (see 
Figure~\ref{dvarhigal}), to estimate an univocal fractal dimension for maps at this wavelength is
particularly difficult and plagued by strong uncertainties. In fact, it is well known that
the $\Delta$-variance is more sensitive to variations of the fractal behavior over different spatial
scale ranges \citep{stu98,oss08a,sch11}, highlighted by distortions or local maxima/minima of the curves, so that a fractal
dimension can be derived in a limited inertial range which is integral part of the information. 
Instead, the curves obtained through box counting algorithms show linearity over larger scale ranges, 
as in Figure~\ref{fdims}, left panel, but the corresponding fractal dimension represents a coarser 
description of the global statistical properties of the images.

%\begin{minipage}{16.0cm}\begin{center}\plotone{./mosaicpaperspire350.eps}\end{center}\end{minipage}



%\bibliography{adssample}
\bibliography{biblio}

\begin{thebibliography}{}
\expandafter\ifx\csname natexlab\endcsname\relax\def\natexlab#1{#1}\fi

\bibitem[{Allan(1966)}]{all66}
Allan, D. 1966, Proceedings of the IEEE, 54, 221

\bibitem[{{Andr{\'e}} {et~al.}(2010){Andr{\'e}}, {Men'shchikov}, {Bontemps},
  {K{\"o}nyves}, {Motte}, {Schneider}, {Didelon}, {Minier}, {Saraceno},
  {Ward-Thompson}, {di Francesco}, {White}, {Molinari}, {Testi}, {Abergel},
  {Griffin}, {Henning}, {Royer}, {Mer{\'{\i}}n}, {Vavrek}, {Attard},
  {Arzoumanian}, {Wilson}, {Ade}, {Aussel}, {Baluteau}, {Benedettini},
  {Bernard}, {Blommaert}, {Cambr{\'e}sy}, {Cox}, {di Giorgio}, {Hargrave},
  {Hennemann}, {Huang}, {Kirk}, {Krause}, {Launhardt}, {Leeks}, {Le Pennec},
  {Li}, {Martin}, {Maury}, {Olofsson}, {Omont}, {Peretto}, {Pezzuto}, {Prusti},
  {Roussel}, {Russeil}, {Sauvage}, {Sibthorpe}, {Sicilia-Aguilar}, {Spinoglio},
  {Waelkens}, {Woodcraft}, \& {Zavagno}}]{and10}
{Andr{\'e}}, P., {Men'shchikov}, A., {Bontemps}, S., {et~al.} 2010, \aap, 518,
  L102

\bibitem[{{Arzoumanian} {et~al.}(2011){Arzoumanian}, {Andr{\'e}}, {Didelon},
  {K{\"o}nyves}, {Schneider}, {Men'shchikov}, {Sousbie}, {Zavagno}, {Bontemps},
  {di Francesco}, {Griffin}, {Hennemann}, {Hill}, {Kirk}, {Martin}, {Minier},
  {Molinari}, {Motte}, {Peretto}, {Pezzuto}, {Spinoglio}, {Ward-Thompson},
  {White}, \& {Wilson}}]{arz11}
{Arzoumanian}, D., {Andr{\'e}}, P., {Didelon}, P., {et~al.} 2011, \aap, 529, L6

\bibitem[{{Bec} \& {Khanin}(2007)}]{bec07}
{Bec}, J., \& {Khanin}, K. 2007, \physrep, 447, 1

\bibitem[{{Bensch} {et~al.}(2001){Bensch}, {Stutzki}, \& {Ossenkopf}}]{ben01}
{Bensch}, F., {Stutzki}, J., \& {Ossenkopf}, V. 2001, \aap, 366, 636

\bibitem[{{Bergin} \& {Tafalla}(2007)}]{ber07}
{Bergin}, E.~A., \& {Tafalla}, M. 2007, \araa, 45, 339

\bibitem[{{Bernard} {et~al.}(2010){Bernard}, {Paradis}, {Marshall}, {Montier},
  {Lagache}, {Paladini}, {Veneziani}, {Brunt}, {Mottram}, {Martin},
  {Ristorcelli}, {Noriega-Crespo}, {Compi{\`e}gne}, {Flagey}, {Anderson},
  {Popescu}, {Tuffs}, {Reach}, {White}, {Benedetti}, {Calzoletti}, {Digiorgio},
  {Faustini}, {Juvela}, {Joblin}, {Joncas}, {Mivilles-Deschenes}, {Olmi},
  {Traficante}, {Piacentini}, {Zavagno}, \& {Molinari}}]{ber10}
{Bernard}, J.-P., {Paradis}, D., {Marshall}, D.~J., {et~al.} 2010, \aap, 518,
  L88

\bibitem[{{Biskamp}(2003)}]{bis03}
{Biskamp}, D. 2003, {Magnetohydrodynamic Turbulence}

\bibitem[{{Boldyrev}(2002)}]{bol02}
{Boldyrev}, S. 2002, \apj, 569, 841

\bibitem[{{Brunt} {et~al.}(2010){Brunt}, {Federrath}, \& {Price}}]{bru10}
{Brunt}, C.~M., {Federrath}, C., \& {Price}, D.~J. 2010, \mnras, 405, L56

\bibitem[{{Campeggio} {et~al.}(2005){Campeggio}, {Elia}, {Maiolo}, {Strafella},
  \& {Cecchi-Pestellini}}]{cam05}
{Campeggio}, L., {Elia}, D., {Maiolo}, B.~M.~T., {Strafella}, F., \&
  {Cecchi-Pestellini}, C. 2005, Journal of Physics Conference Series, 6, 172

\bibitem[{{Campeggio} {et~al.}(2004){Campeggio}, {Strafella}, {Elia}, {Maiolo},
  {Cecchi-Pestellini}, {Aiello}, \& {Pezzuto}}]{cam04}
{Campeggio}, L., {Strafella}, F., {Elia}, D., {et~al.} 2004, \apj, 616, 319

\bibitem[{{Campeggio} {et~al.}(2007){Campeggio}, {Strafella}, {Maiolo}, {Elia},
  \& {Aiello}}]{cam07}
{Campeggio}, L., {Strafella}, F., {Maiolo}, B., {Elia}, D., \& {Aiello}, S.
  2007, \apj, 668, 316

\bibitem[{{Cartwright} {et~al.}(2006){Cartwright}, {Whitworth}, \&
  {Nutter}}]{car06}
{Cartwright}, A., {Whitworth}, A.~P., \& {Nutter}, D. 2006, \mnras, 369, 1411

\bibitem[{{Chappell} \& {Scalo}(2001)}]{cha01}
{Chappell}, D., \& {Scalo}, J. 2001, \apj, 551, 712

\bibitem[{{Compi{\`e}gne} {et~al.}(2010){Compi{\`e}gne}, {Flagey},
  {Noriega-Crespo}, {Martin}, {Bernard}, {Paladini}, \& {Molinari}}]{com10}
{Compi{\`e}gne}, M., {Flagey}, N., {Noriega-Crespo}, A., {et~al.} 2010, \apjl,
  724, L44

\bibitem[{{Elia} {et~al.}(2010){Elia}, {Schisano}, {Molinari}, {Robitaille},
  {Angl{\'e}s-Alc{\'a}zar}, {Bally}, {Battersby}, {Benedettini}, {Billot},
  {Calzoletti}, {di Giorgio}, {Faustini}, {Li}, {Martin}, {Morgan}, {Motte},
  {Mottram}, {Natoli}, {Olmi}, {Paladini}, {Piacentini}, {Pestalozzi},
  {Pezzuto}, {Polychroni}, {Smith}, {Strafella}, {Stringfellow}, {Testi},
  {Thompson}, {Traficante}, \& {Veneziani}}]{eli10}
{Elia}, D., {Schisano}, E., {Molinari}, S., {et~al.} 2010, \aap, 518, L97

\bibitem[{{Elia} {et~al.}(2013){Elia}, {Molinari}, {Fukui}, {Schisano}, {Olmi},
  {Veneziani}, {Hayakawa}, {Pestalozzi}, {Schneider}, {Benedettini}, {di
  Giorgio}, {Ikhenaode}, {Mizuno}, {Onishi}, {Pezzuto}, {Piazzo}, {Polychroni},
  {Rygl}, {Yamamoto}, \& {Maruccia}}]{eli13}
{Elia}, D., {Molinari}, S., {Fukui}, Y., {et~al.} 2013, \apj, 772, 45

\bibitem[{{Elmegreen}(1997{\natexlab{a}})}]{elm97b}
{Elmegreen}, B.~G. 1997{\natexlab{a}}, \apj, 480, 674

\bibitem[{{Elmegreen}(1997{\natexlab{b}})}]{elm97a}
---. 1997{\natexlab{b}}, \apj, 477, 196

\bibitem[{{Elmegreen}(2002)}]{elm02}
---. 2002, \apj, 564, 773

\bibitem[{{Elmegreen} \& {Falgarone}(1996)}]{elm96}
{Elmegreen}, B.~G., \& {Falgarone}, E. 1996, \apj, 471, 816

\bibitem[{{Elmegreen} \& {Scalo}(2004)}]{elm04}
{Elmegreen}, B.~G., \& {Scalo}, J. 2004, \araa, 42, 211

\bibitem[{{Falgarone} {et~al.}(1994){Falgarone}, {Lis}, {Phillips}, {Pouquet},
  {Porter}, \& {Woodward}}]{fal94}
{Falgarone}, E., {Lis}, D.~C., {Phillips}, T.~G., {et~al.} 1994, \apj, 436, 728

\bibitem[{{Falgarone} \& {Phillips}(1996)}]{fal96}
{Falgarone}, E., \& {Phillips}, T.~G. 1996, \apj, 472, 191

\bibitem[{{Falgarone} {et~al.}(1991){Falgarone}, {Phillips}, \&
  {Walker}}]{fal91}
{Falgarone}, E., {Phillips}, T.~G., \& {Walker}, C.~K. 1991, \apj, 378, 186

\bibitem[{{Federrath} {et~al.}(2009){Federrath}, {Klessen}, \&
  {Schmidt}}]{fed09}
{Federrath}, C., {Klessen}, R.~S., \& {Schmidt}, W. 2009, \apj, 692, 364

\bibitem[{{Federrath} {et~al.}(2010){Federrath}, {Roman-Duval}, {Klessen},
  {Schmidt}, \& {Mac Low}}]{fed10}
{Federrath}, C., {Roman-Duval}, J., {Klessen}, R.~S., {Schmidt}, W., \& {Mac
  Low}, M.-M. 2010, \aap, 512, A81

\bibitem[{{Gazol} \& {Kim}(2010)}]{gaz10}
{Gazol}, A., \& {Kim}, J. 2010, \apj, 723, 482

\bibitem[{{Gazol} {et~al.}(2005){Gazol}, {V{\'a}zquez-Semadeni}, \&
  {Kim}}]{gaz05}
{Gazol}, A., {V{\'a}zquez-Semadeni}, E., \& {Kim}, J. 2005, \apj, 630, 911

\bibitem[{{Giannini} {et~al.}(2012){Giannini}, {Elia}, {Lorenzetti},
  {Molinari}, {Motte}, {Schisano}, {Pezzuto}, {Pestalozzi}, {di Giorgio},
  {Andr{\'e}}, {Hill}, {Benedettini}, {Bontemps}, {di Francesco}, {Fallscheer},
  {Hennemann}, {Kirk}, {Minier}, {Nguyen Luong}, {Polychroni}, {Rygl},
  {Saraceno}, {Schneider}, {Spinoglio}, {Testi}, {Ward-Thompson}, \&
  {White}}]{gia12}
{Giannini}, T., {Elia}, D., {Lorenzetti}, D., {et~al.} 2012, \aap, 539, A156

\bibitem[{{Goodman} {et~al.}(2009){Goodman}, {Pineda}, \& {Schnee}}]{goo09}
{Goodman}, A.~A., {Pineda}, J.~E., \& {Schnee}, S.~L. 2009, \apj, 692, 91

\bibitem[{{Griffin} {et~al.}(2010){Griffin}, {Abergel}, {Abreu}, {Ade},
  {Andr{\'e}}, {Augueres}, {Babbedge}, {Bae}, {Baillie}, {Baluteau}, {Barlow},
  {Bendo}, {Benielli}, {Bock}, {Bonhomme}, {Brisbin}, {Brockley-Blatt},
  {Caldwell}, {Cara}, {Castro-Rodriguez}, {Cerulli}, {Chanial}, {Chen},
  {Clark}, {Clements}, {Clerc}, {Coker}, {Communal}, {Conversi}, {Cox},
  {Crumb}, {Cunningham}, {Daly}, {Davis}, {de Antoni}, {Delderfield}, {Devin},
  {di Giorgio}, {Didschuns}, {Dohlen}, {Donati}, {Dowell}, {Dowell}, {Duband},
  {Dumaye}, {Emery}, {Ferlet}, {Ferrand}, {Fontignie}, {Fox}, {Franceschini},
  {Frerking}, {Fulton}, {Garcia}, {Gastaud}, {Gear}, {Glenn}, {Goizel},
  {Griffin}, {Grundy}, {Guest}, {Guillemet}, {Hargrave}, {Harwit}, {Hastings},
  {Hatziminaoglou}, {Herman}, {Hinde}, {Hristov}, {Huang}, {Imhof}, {Isaak},
  {Israelsson}, {Ivison}, {Jennings}, {Kiernan}, {King}, {Lange}, {Latter},
  {Laurent}, {Laurent}, {Leeks}, {Lellouch}, {Levenson}, {Li}, {Li},
  {Lilienthal}, {Lim}, {Liu}, {Lu}, {Madden}, {Mainetti}, {Marliani}, {McKay},
  {Mercier}, {Molinari}, {Morris}, {Moseley}, {Mulder}, {Mur}, {Naylor},
  {Nguyen}, {O'Halloran}, {Oliver}, {Olofsson}, {Olofsson}, {Orfei}, {Page},
  {Pain}, {Panuzzo}, {Papageorgiou}, {Parks}, {Parr-Burman}, {Pearce},
  {Pearson}, {P{\'e}rez-Fournon}, {Pinsard}, {Pisano}, {Podosek}, {Pohlen},
  {Polehampton}, {Pouliquen}, {Rigopoulou}, {Rizzo}, {Roseboom}, {Roussel},
  {Rowan-Robinson}, {Rownd}, {Saraceno}, {Sauvage}, {Savage}, {Savini},
  {Sawyer}, {Scharmberg}, {Schmitt}, {Schneider}, {Schulz}, {Schwartz},
  {Shafer}, {Shupe}, {Sibthorpe}, {Sidher}, {Smith}, {Smith}, {Smith},
  {Spencer}, {Stobie}, {Sudiwala}, {Sukhatme}, {Surace}, {Stevens}, {Swinyard},
  {Trichas}, {Tourette}, {Triou}, {Tseng}, {Tucker}, {Turner}, {Vaccari},
  {Valtchanov}, {Vigroux}, {Virique}, {Voellmer}, {Walker}, {Ward}, {Waskett},
  {Weilert}, {Wesson}, {White}, {Whitehouse}, {Wilson}, {Winter}, {Woodcraft},
  {Wright}, {Xu}, {Zavagno}, {Zemcov}, {Zhang}, \& {Zonca}}]{gri10}
{Griffin}, M.~J., {Abergel}, A., {Abreu}, A., {et~al.} 2010, \aap, 518, L3

\bibitem[{{Gustafsson} {et~al.}(2006){Gustafsson}, {Lemaire}, \&
  {Field}}]{gus06}
{Gustafsson}, M., {Lemaire}, J.~L., \& {Field}, D. 2006, \aap, 456, 171

\bibitem[{{Heithausen} {et~al.}(1998){Heithausen}, {Bensch}, {Stutzki},
  {Falgarone}, \& {Panis}}]{hei98}
{Heithausen}, A., {Bensch}, F., {Stutzki}, J., {Falgarone}, E., \& {Panis},
  J.~F. 1998, \aap, 331, L65

\bibitem[{{Henriksen} \& {Turner}(1984)}]{hen84}
{Henriksen}, R.~N., \& {Turner}, B.~E. 1984, \apj, 287, 200

\bibitem[{{Houlahan} \& {Scalo}(1992)}]{hou92}
{Houlahan}, P., \& {Scalo}, J. 1992, \apj, 393, 172

\bibitem[{{Ingalls} {et~al.}(2004){Ingalls}, {Miville-Desch{\^e}nes}, {Reach},
  {Noriega-Crespo}, {Carey}, {Boulanger}, {Stolovy}, {Padgett}, {Burgdorf},
  {Fajardo-Acosta}, {Glaccum}, {Helou}, {Hoard}, {Karr}, {O'Linger}, {Rebull},
  {Rho}, {Stauffer}, \& {Wachter}}]{ing04}
{Ingalls}, J.~G., {Miville-Desch{\^e}nes}, M.-A., {Reach}, W.~T., {et~al.}
  2004, \apjs, 154, 281

\bibitem[{{Khalil} {et~al.}(2006){Khalil}, {Joncas}, {Nekka}, {Kestener}, \&
  {Arneodo}}]{kha06}
{Khalil}, A., {Joncas}, G., {Nekka}, F., {Kestener}, P., \& {Arneodo}, A. 2006,
  \apjs, 165, 512

\bibitem[{{Kolmogorov}(1941)}]{kol41}
{Kolmogorov}, A. 1941, Akademiia Nauk SSSR Doklady, 30, 301

\bibitem[{{Kowal} {et~al.}(2007){Kowal}, {Lazarian}, \& {Beresnyak}}]{kow07}
{Kowal}, G., {Lazarian}, A., \& {Beresnyak}, A. 2007, \apj, 658, 423

\bibitem[{{Kramer} {et~al.}(1998){Kramer}, {Stutzki}, {Rohrig}, \&
  {Corneliussen}}]{kra98}
{Kramer}, C., {Stutzki}, J., {Rohrig}, R., \& {Corneliussen}, U. 1998, \aap,
  329, 249

\bibitem[{{Kritsuk} \& {Norman}(2004)}]{kri04}
{Kritsuk}, A.~G., \& {Norman}, M.~L. 2004, \apjl, 601, L55

\bibitem[{{Kritsuk} {et~al.}(2006){Kritsuk}, {Norman}, \& {Padoan}}]{kri06}
{Kritsuk}, A.~G., {Norman}, M.~L., \& {Padoan}, P. 2006, \apjl, 638, L25

\bibitem[{{Mandelbrot}(1983)}]{man83}
{Mandelbrot}, B.~B. 1983, {The fractal geometry of nature /Revised and enlarged
  edition/}

\bibitem[{{Martin} {et~al.}(2010){Martin}, {Miville-Desch{\^e}nes}, {Roy},
  {Bernard}, {Molinari}, {Billot}, {Brunt}, {Calzoletti}, {Digiorgio}, {Elia},
  {Faustini}, {Joncas}, {Mottram}, {Natoli}, {Noriega-Crespo}, {Paladini},
  {Robitaille}, {Strafella}, {Traficante}, \& {Veneziani}}]{mar10}
{Martin}, P.~G., {Miville-Desch{\^e}nes}, M.-A., {Roy}, A., {et~al.} 2010,
  \aap, 518, L105

\bibitem[{{Miville-Desch{\^e}nes} {et~al.}(2007){Miville-Desch{\^e}nes},
  {Lagache}, {Boulanger}, \& {Puget}}]{miv07}
{Miville-Desch{\^e}nes}, M.-A., {Lagache}, G., {Boulanger}, F., \& {Puget},
  J.-L. 2007, \aap, 469, 595

\bibitem[{{Miville-Desch{\^e}nes} {et~al.}(2003){Miville-Desch{\^e}nes},
  {Levrier}, \& {Falgarone}}]{miv03}
{Miville-Desch{\^e}nes}, M.-A., {Levrier}, F., \& {Falgarone}, E. 2003, \apj,
  593, 831

\bibitem[{{Mizuno} \& {Fukui}(2004)}]{miz04}
{Mizuno}, A., \& {Fukui}, Y. 2004, in Astronomical Society of the Pacific
  Conference Series, Vol. 317, Milky Way Surveys: The Structure and Evolution
  of our Galaxy, ed. D.~{Clemens}, R.~{Shah}, \& T.~{Brainerd}, 59

\bibitem[{{Molinari} {et~al.}(2010{\natexlab{a}}){Molinari}, {Swinyard},
  {Bally}, {Barlow}, {Bernard}, {Martin}, {Moore}, {Noriega-Crespo}, {Plume},
  {Testi}, {Zavagno}, {Abergel}, {Ali}, {Anderson}, {Andr{\'e}}, {Baluteau},
  {Battersby}, {Beltr{\'a}n}, {Benedettini}, {Billot}, {Blommaert}, {Bontemps},
  {Boulanger}, {Brand}, {Brunt}, {Burton}, {Calzoletti}, {Carey}, {Caselli},
  {Cesaroni}, {Cernicharo}, {Chakrabarti}, {Chrysostomou}, {Cohen},
  {Compiegne}, {de Bernardis}, {de Gasperis}, {di Giorgio}, {Elia}, {Faustini},
  {Flagey}, {Fukui}, {Fuller}, {Ganga}, {Garcia-Lario}, {Glenn}, {Goldsmith},
  {Griffin}, {Hoare}, {Huang}, {Ikhenaode}, {Joblin}, {Joncas}, {Juvela},
  {Kirk}, {Lagache}, {Li}, {Lim}, {Lord}, {Marengo}, {Marshall}, {Masi},
  {Massi}, {Matsuura}, {Minier}, {Miville-Desch{\^e}nes}, {Montier}, {Morgan},
  {Motte}, {Mottram}, {M{\"u}ller}, {Natoli}, {Neves}, {Olmi}, {Paladini},
  {Paradis}, {Parsons}, {Peretto}, {Pestalozzi}, {Pezzuto}, {Piacentini},
  {Piazzo}, {Polychroni}, {Pomar{\`e}s}, {Popescu}, {Reach}, {Ristorcelli},
  {Robitaille}, {Robitaille}, {Rod{\'o}n}, {Roy}, {Royer}, {Russeil},
  {Saraceno}, {Sauvage}, {Schilke}, {Schisano}, {Schneider}, {Schuller},
  {Schulz}, {Sibthorpe}, {Smith}, {Smith}, {Spinoglio}, {Stamatellos},
  {Strafella}, {Stringfellow}, {Sturm}, {Taylor}, {Thompson}, {Traficante},
  {Tuffs}, {Umana}, {Valenziano}, {Vavrek}, {Veneziani}, {Viti}, {Waelkens},
  {Ward-Thompson}, {White}, {Wilcock}, {Wyrowski}, {Yorke}, \&
  {Zhang}}]{mol10b}
{Molinari}, S., {Swinyard}, B., {Bally}, J., {et~al.} 2010{\natexlab{a}}, \aap,
  518, L100

\bibitem[{{Molinari} {et~al.}(2010{\natexlab{b}}){Molinari}, {Swinyard},
  {Bally}, {Barlow}, {Bernard}, {Martin}, {Moore}, {Noriega-Crespo}, {Plume},
  {Testi}, {Zavagno}, {Abergel}, {Ali}, {Andr{\'e}}, {Baluteau}, {Benedettini},
  {Bern{\'e}}, {Billot}, {Blommaert}, {Bontemps}, {Boulanger}, {Brand},
  {Brunt}, {Burton}, {Campeggio}, {Carey}, {Caselli}, {Cesaroni}, {Cernicharo},
  {Chakrabarti}, {Chrysostomou}, {Codella}, {Cohen}, {Compiegne}, {Davis}, {de
  Bernardis}, {de Gasperis}, {Di Francesco}, {di Giorgio}, {Elia}, {Faustini},
  {Fischera}, {Fukui}, {Fuller}, {Ganga}, {Garcia-Lario}, {Giard}, {Giardino},
  {Glenn}, {Goldsmith}, {Griffin}, {Hoare}, {Huang}, {Jiang}, {Joblin},
  {Joncas}, {Juvela}, {Kirk}, {Lagache}, {Li}, {Lim}, {Lord}, {Lucas},
  {Maiolo}, {Marengo}, {Marshall}, {Masi}, {Massi}, {Matsuura}, {Meny},
  {Minier}, {Miville-Desch{\^e}nes}, {Montier}, {Motte}, {M{\"u}ller},
  {Natoli}, {Neves}, {Olmi}, {Paladini}, {Paradis}, {Pestalozzi}, {Pezzuto},
  {Piacentini}, {Pomar{\`e}s}, {Popescu}, {Reach}, {Richer}, {Ristorcelli},
  {Roy}, {Royer}, {Russeil}, {Saraceno}, {Sauvage}, {Schilke},
  {Schneider-Bontemps}, {Schuller}, {Schultz}, {Shepherd}, {Sibthorpe},
  {Smith}, {Smith}, {Spinoglio}, {Stamatellos}, {Strafella}, {Stringfellow},
  {Sturm}, {Taylor}, {Thompson}, {Tuffs}, {Umana}, {Valenziano}, {Vavrek},
  {Viti}, {Waelkens}, {Ward-Thompson}, {White}, {Wyrowski}, {Yorke}, \&
  {Zhang}}]{mol10a}
---. 2010{\natexlab{b}}, \pasp, 122, 314

\bibitem[{{Olmi} {et~al.}(2013){Olmi}, {Angl{\'e}s-Alc{\'a}zar}, {Elia},
  {Molinari}, {Montier}, {Pestalozzi}, {Pezzuto}, {Polychroni}, {Ristorcelli},
  {Rodon}, {Schisano}, {Smith}, {Testi}, \& {Thompson}}]{olm13}
{Olmi}, L., {Angl{\'e}s-Alc{\'a}zar}, D., {Elia}, D., {et~al.} 2013, \aap, 551,
  A111

\bibitem[{{Ossenkopf} {et~al.}(2006){Ossenkopf}, {Esquivel}, {Lazarian}, \&
  {Stutzki}}]{oss06}
{Ossenkopf}, V., {Esquivel}, A., {Lazarian}, A., \& {Stutzki}, J. 2006, \aap,
  452, 223

\bibitem[{{Ossenkopf} {et~al.}(2001){Ossenkopf}, {Klessen}, \&
  {Heitsch}}]{oss01}
{Ossenkopf}, V., {Klessen}, R.~S., \& {Heitsch}, F. 2001, \aap, 379, 1005

\bibitem[{{Ossenkopf} {et~al.}(2008){Ossenkopf}, {Krips}, \&
  {Stutzki}}]{oss08a}
{Ossenkopf}, V., {Krips}, M., \& {Stutzki}, J. 2008, \aap, 485, 917

\bibitem[{{Padoan} {et~al.}(2003){Padoan}, {Boldyrev}, {Langer}, \&
  {Nordlund}}]{pad03}
{Padoan}, P., {Boldyrev}, S., {Langer}, W., \& {Nordlund}, {\AA}. 2003, \apj,
  583, 308

\bibitem[{{Padoan} {et~al.}(2006){Padoan}, {Cambr{\'e}sy}, {Juvela}, {Kritsuk},
  {Langer}, \& {Norman}}]{pad06}
{Padoan}, P., {Cambr{\'e}sy}, L., {Juvela}, M., {et~al.} 2006, \apj, 649, 807

\bibitem[{{Padoan} {et~al.}(2002){Padoan}, {Cambr{\'e}sy}, \&
  {Langer}}]{padc02}
{Padoan}, P., {Cambr{\'e}sy}, L., \& {Langer}, W. 2002, \apjl, 580, L57

\bibitem[{{Padoan} \& {Nordlund}(2002)}]{padn02}
{Padoan}, P., \& {Nordlund}, {\AA}. 2002, \apj, 576, 870

\bibitem[{Peitgen \& Saupe(1988)}]{pei88}
Peitgen, H.-O., \& Saupe, D., eds. 1988, The Science of Fractal Images, xiii +
  312

\bibitem[{{Pilbratt} {et~al.}(2010){Pilbratt}, {Riedinger}, {Passvogel},
  {Crone}, {Doyle}, {Gageur}, {Heras}, {Jewell}, {Metcalfe}, {Ott}, \&
  {Schmidt}}]{pil10}
{Pilbratt}, G.~L., {Riedinger}, J.~R., {Passvogel}, T., {et~al.} 2010, \aap,
  518, L1

\bibitem[{{Poglitsch} {et~al.}(2010){Poglitsch}, {Waelkens}, {Geis},
  {Feuchtgruber}, {Vandenbussche}, {Rodriguez}, {Krause}, {Renotte}, {van
  Hoof}, {Saraceno}, {Cepa}, {Kerschbaum}, {Agn{\`e}se}, {Ali}, {Altieri},
  {Andreani}, {Augueres}, {Balog}, {Barl}, {Bauer}, {Belbachir}, {Benedettini},
  {Billot}, {Boulade}, {Bischof}, {Blommaert}, {Callut}, {Cara}, {Cerulli},
  {Cesarsky}, {Contursi}, {Creten}, {De Meester}, {Doublier}, {Doumayrou},
  {Duband}, {Exter}, {Genzel}, {Gillis}, {Gr{\"o}zinger}, {Henning},
  {Herreros}, {Huygen}, {Inguscio}, {Jakob}, {Jamar}, {Jean}, {de Jong},
  {Katterloher}, {Kiss}, {Klaas}, {Lemke}, {Lutz}, {Madden}, {Marquet},
  {Martignac}, {Mazy}, {Merken}, {Montfort}, {Morbidelli}, {M{\"u}ller},
  {Nielbock}, {Okumura}, {Orfei}, {Ottensamer}, {Pezzuto}, {Popesso},
  {Putzeys}, {Regibo}, {Reveret}, {Royer}, {Sauvage}, {Schreiber}, {Stegmaier},
  {Schmitt}, {Schubert}, {Sturm}, {Thiel}, {Tofani}, {Vavrek}, {Wetzstein},
  {Wieprecht}, \& {Wiezorrek}}]{pog10}
{Poglitsch}, A., {Waelkens}, C., {Geis}, N., {et~al.} 2010, \aap, 518, L2

\bibitem[{{Rosas-Guevara} {et~al.}(2010){Rosas-Guevara},
  {V{\'a}zquez-Semadeni}, {G{\'o}mez}, \& {Jappsen}}]{ros10}
{Rosas-Guevara}, Y., {V{\'a}zquez-Semadeni}, E., {G{\'o}mez}, G.~C., \&
  {Jappsen}, A.-K. 2010, \mnras, 406, 1875

\bibitem[{{Rosner} \& {Bodo}(1996)}]{ros96}
{Rosner}, R., \& {Bodo}, G. 1996, \apjl, 470, L49

\bibitem[{{Rowles} \& {Froebrich}(2011)}]{row11}
{Rowles}, J., \& {Froebrich}, D. 2011, \mnras, 416, 294

\bibitem[{{Russeil} {et~al.}(2013){Russeil}, {Schneider}, {Anderson},
  {Zavagno}, {Molinari}, {Persi}, {Bontemps}, {Motte}, {Ossenkopf},
  {Andr{\'e}}, {Arzoumanian}, {Bernard}, {Deharveng}, {Didelon}, {Di
  Francesco}, {Elia}, {Hennemann}, {Hill}, {K{\"o}nyves}, {Li}, {Martin},
  {Nguyen Luong}, {Peretto}, {Pezzuto}, {Polychroni}, {Roussel}, {Rygl},
  {Spinoglio}, {Testi}, {Tig{\'e}}, {Vavrek}, {Ward-Thompson}, \&
  {White}}]{rus13}
{Russeil}, D., {Schneider}, N., {Anderson}, L.~D., {et~al.} 2013, \aap, 554,
  A42

\bibitem[{{S{\'a}nchez} {et~al.}(2005){S{\'a}nchez}, {Alfaro}, \&
  {P{\'e}rez}}]{san05}
{S{\'a}nchez}, N., {Alfaro}, E.~J., \& {P{\'e}rez}, E. 2005, \apj, 625, 849

\bibitem[{{S{\'a}nchez} {et~al.}(2007){S{\'a}nchez}, {Alfaro}, \&
  {P{\'e}rez}}]{san07}
---. 2007, \apj, 656, 222

\bibitem[{{S{\'a}nchez} {et~al.}(2009){S{\'a}nchez}, {Alfaro}, \&
  {P{\'e}rez}}]{san09}
{S{\'a}nchez}, N., {Alfaro}, E.~J., \& {P{\'e}rez}, E. 2009, in Revista
  Mexicana de Astronomia y Astrofisica, vol. 27, Vol.~35, Revista Mexicana de
  Astronomia y Astrofisica Conference Series, 76--77

\bibitem[{Sarkar \& Chaudhuri(1994)}]{sar94}
Sarkar, N., \& Chaudhuri, B. 1994, Systems, Man and Cybernetics, IEEE
  Transactions on, 24, 115

\bibitem[{{Scalo}(1990)}]{sca90}
{Scalo}, J. 1990, in Astrophysics and Space Science Library, Vol. 162, Physical
  Processes in Fragmentation and Star Formation, ed. R.~{Capuzzo-Dolcetta},
  C.~{Chiosi}, \& A.~{di Fazio}, 151--176

\bibitem[{{Schisano} {et~al.}(2013){Schisano}, {Rygl}, {Molinari}, {Busquet},
  {Elia}, {Pestalozzi}, {Polychroni}, {Billot}, {Glover}, {Noriega-Crespo},
  {Paladini}, {Moore}, {Plume}, \& {V\'{a}zquez-Semadeni}}]{sch13}
{Schisano}, E., {Rygl}, K.~L.~J., {Molinari}, S., {et~al.} 2013, \apj,
  submitted

\bibitem[{{Schneider} {et~al.}(1998){Schneider}, {Stutzki}, {Winnewisser}, \&
  {Block}}]{sch98}
{Schneider}, N., {Stutzki}, J., {Winnewisser}, G., \& {Block}, D. 1998, \aap,
  335, 1049

\bibitem[{{Schneider} {et~al.}(2011){Schneider}, {Bontemps}, {Simon},
  {Ossenkopf}, {Federrath}, {Klessen}, {Motte}, {Andr{\'e}}, {Stutzki}, \&
  {Brunt}}]{sch11}
{Schneider}, N., {Bontemps}, S., {Simon}, R., {et~al.} 2011, \aap, 529, A1

\bibitem[{{Shadmehri} \& {Elmegreen}(2011)}]{sha11}
{Shadmehri}, M., \& {Elmegreen}, B.~G. 2011, \mnras, 410, 788

\bibitem[{{Sharpless}(1959)}]{sha59}
{Sharpless}, S. 1959, \apjs, 4, 257

\bibitem[{{Sreenivasan} {et~al.}(1989){Sreenivasan}, {Ramshankar}, \&
  {Meneveau}}]{sre89}
{Sreenivasan}, K.~R., {Ramshankar}, R., \& {Meneveau}, C. 1989, Royal Society
  of London Proceedings Series A, 421, 79

\bibitem[{{Stutzki} {et~al.}(1998){Stutzki}, {Bensch}, {Heithausen},
  {Ossenkopf}, \& {Zielinsky}}]{stu98}
{Stutzki}, J., {Bensch}, F., {Heithausen}, A., {Ossenkopf}, V., \& {Zielinsky},
  M. 1998, \aap, 336, 697

\bibitem[{{Traficante} {et~al.}(2011){Traficante}, {Calzoletti}, {Veneziani},
  {Ali}, {de Gasperis}, {di Giorgio}, {Faustini}, {Ikhenaode}, {Molinari},
  {Natoli}, {Pestalozzi}, {Pezzuto}, {Piacentini}, {Piazzo}, {Polenta}, \&
  {Schisano}}]{tra11}
{Traficante}, A., {Calzoletti}, L., {Veneziani}, M., {et~al.} 2011, \mnras,
  416, 2932

\bibitem[{{Vavrek} {et~al.}(2001){Vavrek}, {Bal{\'a}zs}, \& {Epchtein}}]{vav01}
{Vavrek}, R., {Bal{\'a}zs}, L.~G., \& {Epchtein}, N. 2001, in Astronomical
  Society of the Pacific Conference Series, Vol. 243, From Darkness to Light:
  Origin and Evolution of Young Stellar Clusters, ed. T.~{Montmerle} \&
  P.~{Andr{\'e}}, 149

\bibitem[{{V{\'a}zquez-Semadeni}(1999)}]{vaz99}
{V{\'a}zquez-Semadeni}, E. 1999, in Astrophysics and Space Science Library,
  Vol. 241, Millimeter-Wave Astronomy: Molecular Chemistry {\&} Physics in
  Space., ed. W.~F. {Wall}, A.~{Carrami{\~n}ana}, \& L.~{Carrasco}, 161

\bibitem[{{Wilson} {et~al.}(1999){Wilson}, {Howe}, \& {Balogh}}]{wil99b}
{Wilson}, C.~D., {Howe}, J.~E., \& {Balogh}, M.~L. 1999, \apj, 517, 174

\end{thebibliography}

\clearpage

%% the file, uncomment the \plotone, \plottwo, and
%% If you need a layout that cannot be achieved with \plotone or
%% \plottwo, you can invoke the graphicx package directly with the
%% \includegraphics command or use \plotfiddle. For more information,
%% please see the tutorial on "Using Electronic Art with AASTeX" in the
%% documentation section at the AASTeX Web site,
%% http://www.journals.uchicago.edu/AAS/AASTeX.
%%
%% The examples below also include sample markup for submission of
%% supplemental electronic materials. As always, be sure to check
%% the instructions to authors for the journal you are submitting to
%% for specific submissions guidelines as they vary from

%% This example uses \plotone to include an EPS file scaled to
%% 80% of its natural size with \epsscale. Its caption
%% has been written to indicate that additional figure parts will be
%% available in the electronic journal.

%\begin{figure}
%\epsscale{.80}
%\plotone{f1.eps}
%\caption{Derived spectra for 3C138 \citep[see][]{heiles03}. Plots for all sources are available
%in the electronic edition of {\it The Astrophysical Journal}.\label{fig1}}
%\end{figure}

\clearpage

\clearpage

\clearpage

%% The following command ends your manuscript. LaTeX will ignore any text

\end{document}